\documentclass[epj,nopacs]{svjour}
\usepackage{slashed}
\usepackage{soul}
\usepackage{mathrsfs,bm}
\usepackage{txfonts}
\usepackage{amssymb}

\usepackage{amsmath}
\usepackage{indentfirst}
\usepackage{graphicx,booktabs}
\usepackage{multirow}
\usepackage{overpic}
\usepackage{color}

\usepackage{hyperref}
\usepackage{doi}
\usepackage{graphics}
\usepackage{epsfig}
\usepackage{soul}
\usepackage{color}

\usepackage[sort&compress,numbers]{natbib}

\usepackage{hyperref}

\usepackage[utf8]{inputenc}

\newcommand{\corr}[1]{#1}

\begin{document}

\title{Interpretations of the new LHCb $P_c(4337)^+$ pentaquark state}

\date{\today}

\author{Mao-Jun Yan \inst{1} \and
  Fang-Zheng Peng \inst{2} \and
  Mario {S\'anchez S\'anchez} \inst{3} \and
  Manuel {Pavon} Valderrama \inst{2} \mail{mpavon@buaa.edu.cn}}

\institute{CAS Key Laboratory of Theoretical Physics, 
  Institute of Theoretical Physics,
  Chinese Academy of Sciences, Beijing 100190, China \and
  School of Physics, Beihang University, Beijing 100191, China \and
  Centre d'\'Etudes Nucl\'eaires, CNRS/IN2P3, Universit\'e de Bordeaux, 33175 Gradignan, France \
}

\abstract{
  Recently the LHCb collaboration has observed a new pentaquark state,
  the $P_c(4337)^+$.
  Owing to its proximity to the $\chi_{c0}(1S) p$, $\bar{D}^* \Lambda_c$,
  $\bar{D} \Sigma_c$ and $\bar{D} \Sigma_c^*$ thresholds,
  this new pentaquark might very well be a meson-baryon
  bound state.
  However its spin and parity have not been determined yet and
  none of the previous possibilities can be ruled out.
  We briefly explore a few of these options and the consequences they entail
  in the present manuscript: (i) the $P_c(4337)^+$ might be a $\chi_{c0}(1S) p$
  bound state, (ii) the $P_c(4312)^+$ and $P_c(4337)^+$
  might be $\bar{D}^* \Lambda_c$
  and $\bar{D} \Sigma_c$ states close to threshold, respectively,
  where the Breit-Wigner
  mass might not correspond to the location of the poles, (iii)
  the locations of the $P_c(4312)^+$ and $P_c(4337)^+$ might be
  explained in terms of the $\bar{D}^* \Lambda_c$-$\bar{D} \Sigma_c$
  and $\bar{D}^* \Lambda_c$-$\bar{D} \Sigma_c^*$ coupled channel
  dynamics.
  This last option, though not the most probable explanation, is still
  potentially compatible with the double peak
  solution of the $P_{cs}(4459)^0$ and with what we know of
  the $P_c(4312)^+$.
  As a byproduct of the previous explorations, we conjecture the existence
  of a series of anticharmed meson - antitriplet charmed baryon
  bound states and calculate their masses.
  }

\maketitle

\section{Introduction}
The LHCb collaboration has announced~\cite{Aaij:2021august}
the observation of a new pentaquark in the $J/\psi p$ invariant mass
distribution, where its mass and width are
\begin{eqnarray}
  M(P_c(4337)^+) &=& 4337^{+7}_{-4} {}^{+2}_{-2} \,  {\rm MeV}  \, , \\
  \Gamma(P_c(4337)^+) &=& 29^{+26}_{-12} {}^{+14}_{-14} \,  {\rm MeV}  \, , 
\end{eqnarray}
and the statistical significance of the signal varies between
$3.1-3.7\,\sigma$ depending on the $J^P$ assignment.
This pentaquark, being relatively narrow, might be related to the already
known $P_c(4312)^+$, $P_c(4440)^+$, $P_c(4457)^+$~\cite{LHCb:2019kea}
and $P_{cs}(4459)^0$~\cite{Aaij:2020gdg} (where from now on we will drop
their charge superscript).
Of course the question is what the nature of this state is: is it a compact
pentaquark state or is it molecular? Does it have partners?
Can it be grouped together with other known pentaquarks? 

Here we will briefly review a few of the possibilities for explaining
this new pentaquark (where we will concentrate on meson-baryon
explanations) and the consequences that they entail.
But before that, we comment on the fact that the $P_c(4312)$ is not observed
in the new LHCb results~\cite{Aaij:2021august}, which is puzzling at first
sight, but might be explained by the different production mechanisms
($\Lambda_b^0 \to J/\psi\,p K^-$ and
$B_s^0 \to J\/\psi\,p\,\bar{p}$), though other explanations
are also possible (e.g. poor statistics,
they might be the same state, etc.).
Thus for most of this manuscript we will assume that both the $P_c(4312)$ and
$P_c(4337)$ exist and that they are different states.
The possibilities we will explore are: (i) the $P_c(4337)$ is a hadrocharmonium,
(ii) the actual masses of the $P_c(4312)$ and $P_c(4337)$ do not coincide
with the experimental ones obtained from their Breit-Wigner
parametrizations, and they are actually $\bar{D}^* \Lambda_c$ and
$\bar{D} \Sigma_c$ molecular states, respectively, and
(iii) the $P_c(4312)$ and $P_c(4337)$ are poles of the
$\bar{D}^* \Lambda_c$-$\bar{D} \Sigma_c$ and
$\bar{D}^* \Lambda_c$-$\bar{D} \Sigma_c^*$
coupled channel systems.
This last option will be particularly interesting, though not for its power
for explaining the $P_c(4337)$. Instead, it allows to deduce the probable
existence of a series of anticharmed meson - antitriplet charmed baryon
pentaquarks and to calculate their masses.
It is worth mentioning that other works have also explored variations of
the previous possibilities or new explanations:
for instance, Ref.~\cite{Nakamura:2021dix} has proposed that
the $P_c(4312)$ and $P_c(4337)$ are simply two
manifestations of the same state,
while Ref.~\cite{Ferretti:2021zis} suggests that the $P_c(4337)$ and
$P_{cs}(4459)$ are actually part of the same hadrocharmonium octet.

\section{Hadrocharmonium}
A very straightforward explanation is that of a $\chi_{c0}(1S) p$ bound
state~\cite{Eides:2015dtr,Eides:2017xnt,Eides:2019tgv},
i.e. a hadrocharmonium state: the $\chi_{c0} p$ threshold is located
merely $10\,{\rm MeV}$ above the $P_c(4337)$.
The binding mechanism is a two-gluon exchange short-range force between
the $\chi_{c0}$ and the proton, the strength of which is proportional to
the chromopolarizability $\alpha_S$ of the charmonium.
Chromopolarizabilities are in general not well known, but large-$N_c$
estimations~\cite{Eides:2017xnt} indicate that it might very well
be strong enough as to bind the $\chi_{c0} p$ and other
charmonium - light baryon systems~\cite{Eides:2019tgv}.
Besides, the now suspected existence of a  $J/\psi J/\psi$ bound
state~\cite{Dong:2020nwy}, which depends on similar ingredients (binding
mechanism, estimations of $\alpha_S$, etc.)~\cite{Dong:2021lkh},
if anything makes the hadrocharmonium explanation more believable.
Though originally proposed to explain the $P_c(4312)$~\cite{Eides:2019tgv},
the uncertainties in $\alpha_S$ are likely to be compatible
with the $P_c(4337)$ being the $\chi_{c0} p$ hadrocharmonium
(at least if binding energies are as dependent
on $\alpha_S$ as in~\cite{Eides:2017xnt,Ferretti:2020ewe};
in this regard we notice that
Ref.~\cite{Ferretti:2021zis} suggests an uncertainty of
up to $100\,{\rm MeV}$ in the mass of the prospective $\chi_{c0} p$ bound state
based on two different fits of $\alpha_S$
).
This explanation will predict $J^P = \frac{1}{2}^+$ (which is incidentally
the most favored $J^P$ by the experimental data~\cite{Aaij:2021august}) and
a decay width of the order of the $\chi_{c0}$ charmonium width,
i.e. about $10\,{\rm MeV}$, to which we could add contributions
from short-range $\chi_{c0} p \to \psi(1S,2S) p$ P-wave operators.
This could be consistent with the experimental decay width of
the $P_c(4337)$ once we take into account the admittedly
large uncertainties.

This explanation will also effectively decouple the $P_c(4337)$ from
the other prospective charmed baryon - charmed antimeson molecular
pentaquarks, as the $\chi_{c0} p \to \bar{D}^{(*)} \Sigma_c^{(*)}$
transitions are also P-wave and short-range (they involve the exchange of
a charmed meson) and thus expected to be suppressed.
If this happens to be the case, the current molecular descriptions of
the $P_c(4312)$, $P_c(4440)$ and $P_c(4457)$ pentaquarks
will be left unchanged.
These descriptions are usually based on the lowest order,
heavy-quark spin symmetric $\bar{D}^{(*)} \Sigma_c^{(*)}$ contact-range
potential~\cite{Liu:2018zzu} and usually lead to
the prediction of the same set of partners~\cite{Liu:2019tjn,Xiao:2019aya,Valderrama:2019chc,Liu:2019zvb,Guo:2019fdo,Du:2019pij,Du:2021fmf}
(with a few differences
if pion dynamics~\cite{Yamaguchi:2019seo,Wang:2019ato} or
the $\bar{D} \Lambda_{c1}(2595)$ channel~\cite{Burns:2015dwa,Geng:2017hxc,Burns:2019iih,Peng:2020gwk} are included, while non-molecular schemes tend to
predict a different set of partners~\cite{Ortiz-Pacheco:2018ccl,Wang:2019got};
it is also worth noticing the possibility that only a subset of the previous
three pentaquarks might be molecular~\cite{Kuang:2020bnk}).

\section{Shallow $\bar{D}^* \Lambda_c$ and $\bar{D} \Sigma_c$ states}
A second, relatively prosaic explanation, involves not only the new $P_c(4337)$
pentaquark but also the previous $P_c(4312)$ as well,
whose mass and width are~\cite{LHCb:2019kea}
\begin{eqnarray}
  M(P_c(4312)) &=& 4311.9 \pm 0.7^{+6.8}_{-0.6} \, {\rm MeV}, \\
  \Gamma(P_c(4312)) &=& 9.8 \pm 2.7^{+3.7}_{-4.5} \, {\rm MeV}.
\end{eqnarray}  
Here it is worth noticing that the difference in the masses of the $P_c(4312)$
and $P_c(4337)$ pentaquarks is
\begin{eqnarray}
  M(P_c(4337)) - M(P_c(4312))) \approx 25\,{\rm MeV} \, ,
\end{eqnarray}
which incidentally coincides with the mass difference between
the $\bar{D}^* \Lambda_c$ and $\bar{D} \Sigma_c$ thresholds
\begin{eqnarray}
  M(\bar{D} \Sigma_c) - M(\bar{D}^* \Lambda_c)  = 25.8\,{\rm MeV}  .
\end{eqnarray}
This is interesting, because if there are shallow $\bar{D}^* \Lambda_c$ and
$\bar{D} \Sigma_c$ bound states, the Breit-Wigner parametrization used to
determine their masses is likely to fail.
For instance, the $Z_c(3900)$ and $Z_{cs}(3985)$ Breit-Wigner masses are above
the $D^* \bar{D}$ and $D^* \bar{D}_s$-$D \bar{D}_s^*$ thresholds,
yet theoretical analyses indicate that they might very well
correspond with poles in $D^* \bar{D}$ and $D^* \bar{D}_s$-$D \bar{D}_s^*$
scattering below threshold~\cite{Albaladejo:2015lob,Yang:2020nrt}.
Regarding the $P_c(4312)$ the analysis of the JPAC collaboration already
contemplated the possibility that it is a $\bar{D} \Sigma_c$ virtual
state just below threshold, instead of the more common interpretation
as a $\bar{D} \Sigma_c$ bound state at about $9\,{\rm MeV}$ below
threshold~\cite{Fernandez-Ramirez:2019koa}.
We notice that the JPAC analysis~\cite{Fernandez-Ramirez:2019koa}
did not include the $\bar{D}^* \Lambda_c$ channel,
which might alter their previous conclusions.

The most interesting consequence of this scenario would be the existence
of $\bar{D} \Lambda_c$ and $\bar{D}^* \Lambda_c$ pentaquarks,
which have indeed been theorized~\cite{Chen:2017vai}.
From a theoretical viewpoint their existence is precarious:
in the one-boson-exchange (OBE) model the $\bar{D}^{(*)}$
and $\Lambda_c$ can exchange two light-mesons,
the sigma and the omega~\footnote{The reason for discarding other light mesons
  is that the exchange of a pion is precluded by isospin symmetry and HQSS,
  while rho exchange is forbidden only by isospin.},
where the contribution of the former is attractive and the later repulsive,
leading to a potential of the type~\cite{Chen:2017vai}
\begin{eqnarray}
  V_{\rm OBE}(\bar{D}^* \Lambda_c) &=& \frac{g_{\omega 1} g_{\omega 2}}{4 \pi}\,
  \frac{e^{-m_{\omega} r}}{r} - \frac{g_{\sigma 1} g_{\sigma 2}}{4 \pi}\,
  \frac{e^{-m_{\sigma} r}}{r} \, , \label{eq:VOBE-DLambdac}
\end{eqnarray}
where the indices $i = 1, 2$ refer to the charmed antimeson and charmed baryon,
respectively, $g_{\omega i}$ and $g_{\sigma i}$ are the couplings to the omega
and sigma, and $m_{\omega} = 780\,{\rm MeV}$ and
$m_{\sigma} \sim (400-600)\,{\rm MeV}$ are
the omega and sigma masses.
Phenomenologically (vector meson dominance and linear sigma model) we expect
$g_{\omega 1} = g_{\omega 2} / 2 = m_V / ({2} f_{\pi}) \sim 2.9$ and
$g_{\sigma 1} = g_{\sigma 2} / 2 = \sqrt{2}\,m_N / (3 f_{\pi}) \sim 3.4$,
with $m_V$, $m_N$ the vector meson and nucleon masses and
$f_{\pi} = 130\,{\rm MeV}$ the pion decay constant.
The previous estimation of the couplings will give the upper hand to
the attraction provided by the sigma meson ($g_{\sigma i} > g_{\omega i}$),
potentially leading to bound states, as has been already
proposed~\cite{Chen:2017vai}.
However, considering the approximate nature of the previous relations
this is far from an established conclusion and the interaction
could in principle be repulsive instead.
In this regard, Ref.~\cite{Shen:2017ayv} also points out towards a possible
$\bar{D} \Lambda_c$ bound state, while Refs.~\cite{He:2019rva,Ke:2019bkf}
are less optimistic about this prospect.
Thus determining whether one of the two $P_c(4312)$ and $P_c(4337)$ pentaquarks
could correspond to a $\bar{D}^* \Lambda_c$ bound state will indeed have
important consequences regarding the OBE model
as applied to heavy hadrons.

From the effective field theory (EFT) perspective,
the consequences of this scenario are also interesting.
In EFT the charmed antimeson - charmed baryon interaction can be described
with a non-relativistic potential that admits a low energy power-series
expansion in terms of the ratio $Q/M$, where $Q$ represents a low-energy
scale (e.g. the pion mass or the momentum of the charmed hadrons) and
$M$ a high-energy scale (e.g. the rho mass or the momentum for which
an external probe will be able to discern the internal structure of
the charmed hadrons).
Actually, in a theory with bound states (e.g. molecular pentaquarks)
where pion exchanges and coupled channel effects are perturbative,
the EFT potential at lowest or leading order (${\rm LO}$)
takes the form of a momentum- and energy-independent
contact-range potential
\begin{eqnarray}
  V_C(\vec{q}) = c \, ,
\end{eqnarray}
with $c$ a coupling constant {representing the physics from the degrees of
  freedom not explicitly included in the EFT, e.g. scalar and vector meson
  exchanges, or from the high momentum modes of the degrees of freedom
  already included (i.e. renormalization, where we will elaborate
  later). This coupling}
can be further decomposed into different
contributions depending on the quantum numbers and symmetries of
the system.
If we consider $\bar{D}^{(*)} \Lambda_c$, this is a particular instance of a
charmed antimeson - antitriplet charmed baryon system,
for which the potential reads
\begin{eqnarray}
  V_C(\bar{H}_c T_c) = \lambda^{(S)} d_a^{(S)} + \lambda^{(O)} d_a^{(O)}  \, ,
\end{eqnarray}
with $\bar{H}_c = \bar{D}$, $\bar{D}_s$ or $\bar{D}^*$, $\bar{D}_s^*$,
$T_c = \Lambda_c$, $\Xi_c$, where we have expanded in terms of
SU(3)-flavor representations, with $d_a^{(S)}$ and $d_a^{(O)}$ the singlet
and octet couplings and $\lambda^{(S)}$ and $\lambda^{(O)}$ coefficients.
We notice that there is no spin dependence as this is forbidden
by heavy-quark spin symmetry (HQSS) (the reason being that the total
light spin of the light diquark within the antitriplet baryons
is zero).

If we particularize for the $\bar{D}^* \Lambda_c$ and the isoscalar $I=0$
$\bar{D}^{(*)} \Xi_c$ systems, we find
\begin{eqnarray}
  V_C (\bar{D}^{(*)} \Lambda_c) &=& d_a^{(O)} = \tilde{d}_a \, ,
  \label{eq:Vc-DLambdac} \\
  V_C (\bar{D}^{(*)} \Xi_c, I = 0) &=& \frac{2}{3}\,d_a^{(S)} +
  \frac{1}{3}d_a^{(O)}
  = {d}_a \, , \label{eq:V-DXic}
\end{eqnarray}
where for notational convenience we have defined the couplings $\tilde{d}_a$
and $d_a$.
The reason why we bring up the $\bar{D}^{(*)} \Xi_c$ system is
the recent observation of the $P_{cs}(4459)$
pentaquark~\cite{Aaij:2020gdg}, the mass and width of which are
\begin{eqnarray}
  M_{P_{cs}} = 4458.8 \pm 2.9 {}^{+4.7}_{-1.1} \,{\rm MeV} \, , \nonumber \\
  \Gamma_{P_{cs}} = 17.3 \pm 6.5 {}^{+8.0}_{-5.7} \,{\rm MeV}\, ,
  \label{eq:Pcs-single}
\end{eqnarray}
where its most usual molecular interpretation is that of a $\bar{D}^* \Xi_c$
bound state with
$I=0$~\cite{Chen:2020uif,Liu:2020hcv,Peng:2020hql,Dong:2021juy},
an interpretation further supported by previous predictions of
$\bar{D}^* \Xi_c$ states~\cite{Xiao:2019gjd,Wang:2019nvm}
with similar masses (and widths in the case of~\cite{Xiao:2019gjd}).
This system depends on the $d_a$ coupling defined above, which should be
attractive and strong enough as to bind the system
if the $P_{cs}(4459)$ is indeed a molecular state.
Now, if the $P_c(4312)$ and $P_c(4337)$ were to be $\bar{D}^* \Lambda_c$ and
$\bar{D} \Sigma_c$ shallow bound or virtual states, this will in turn imply
an attractive $\tilde{d}_a$ coupling (though probably not as strong as $d_a$).
From SU(3)-flavor symmetry the interaction of the other S-wave $\bar{H}_c T_c$
systems can be expressed in terms of $\tilde{d}_a$ and $d_a$
\begin{eqnarray}
  V_C (\bar{D}_s^{(*)} \Lambda_c) &=&
 \frac{1}{2} ( {d}_a + \tilde{d}_a) \, , \label{eq:Vc-DsLambdac} \\
  V_C (\bar{D}^{(*)} \Xi_c, I = 1) &=& \tilde{d}_a \, , \\
  V_C (\bar{D}_s^{(*)} \Xi_c) &=& \tilde{d}_a \label{eq:Vc-DsXic} \, .
\end{eqnarray}
Provided that the $\bar{D}^* \Lambda_c$ and $I=0$ $\bar{D} \Xi_c^*$ bind
or are close to binding, this will imply that all the previous systems are
also attractive enough as to generate bound or virtual states near
threshold, which will be possible to detect in the $J/\psi\,\Lambda$,
$J/\psi\,\Sigma$ and $J/\psi\,\Xi$ channels.

\section{$\bar{D}^* \Lambda_c$-$\bar{D} \Sigma_c$ and $\bar{D}^* \Lambda_c$-$\bar{D} \Sigma_c^*$ states}
\label{sec:EFT-A}

We now consider the third interpretation, namely
that the $P_c(4312)$ and $P_c(4337)$ are poles within
the $\bar{D}^* \Lambda_c$-$\bar{D} \Sigma_c$ and
$\bar{D}^* \Lambda_c$-$\bar{D} \Sigma_c^*$
coupled channel dynamics.
This idea, though interesting, will fail to provide a unified description of
the $P_c(4312)$ and $P_c(4337)$ pentaquarks, at least within
the uncertainties of the EFT we will be using.
What we will get instead is the prediction of a $\bar{D} \Sigma_c^*$ pentaquark
in the vicinity of $4370\,{\rm MeV}$, in agreement with 
previous theoretical
works~\cite{Liu:2019tjn,Liu:2019zvb,Guo:2019fdo,Du:2019pij,Du:2021fmf}.

From a theoretical perspective the previous coupled channel dynamics are
really interesting, as they raise the question of how to incorporate
them within the EFT formalism.
This is done by proposing a {\it power counting}, i.e. a principle
by which to order the EFT contributions to the meson-baryon
potential from more to less relevant at low energies.
In this case the EFT description will require more elaboration
than the single channel contact-range potential
we discussed in the previous section.
Before presenting this description in detail, we provide a brief overview of
a few relevant EFT ideas in the following lines.

The interpretation of pentaquark states as shallow
meson-baryon bound states allows the application of the EFT formalism.
The reason is that
the molecular picture implicitly assumes the existence of a separation of
scales, as this description only makes sense when the size of a pentaquark
as a bound system is larger than the size of its components.
EFTs are constructed by writing down all possible interactions involving
the low energy degrees of freedom of the theory (in our case charmed
hadrons and pions) that are compatible with the known low energy
symmetries (most notably HQSS, flavor and chiral symmetries).
In principle this generates an infinite number of interactions and couplings,
though not all them are equally important.
Indeed, EFT interactions can be ordered from more to less relevant at
low energies by means of a power counting, a criterion by which to
decide what is the size of a given coupling.

It is important to emphasize that power counting is not uniquely determined
within a given EFT, but instead depends on choices regarding
the expected size of several physical effects.
Owing to a scarcity of experimental data that can directly constrain
the meson-baryon interactions, the determination of the power
counting will rely on a series of assumptions.
At this point two warnings are worth mentioning:
the first is that changing these assumptions changes the counting.
The second is that 
even for systems for which experimental data are abundant,
such as the two-nucleon system, completely reasonable power counting
expectations have been subverted upon closer theoretical examination,
a very well-known example being how
the KSW counting~\cite{Kaplan:1998tg,Kaplan:1998we} was discovered
not to generate a convergent EFT expansion~\cite{Fleming:1999ee}.

In this manuscript we will explore power counting in a constructive manner:
first, we state a series of assumptions and the consequences they entail,
including what type of pentaquark spectrum they predict.
Then, we will revisit the assumptions and refine them, leading to
modifications of the power counting and new predictions.
We will end up with three power countings ($A$, $B$ and $C$).
In this section we deal with the first of these countings, $A$,
which we determine by following these steps:
\begin{itemize}
\item[a)] We begin by discussing the general properties of
  the EFT describing a two-hadron bound state, including
  the counting of the contact-range potential
  and the pions (if present).

\item[b)] We apply the previous ideas to determine the counting of
  the single channel $\bar{D}^* \Lambda_c$ and $\bar{D} \Sigma_c^{(*)}$ systems
  at low energies.

\item[c)] Then we consider the role of
  the $\bar{D}^* \Lambda_c$-$\bar{D} \Sigma_c$ and
  $\bar{D}^* \Lambda_c$-$\bar{D} \Sigma_c^*$ coupled channel dynamics,
  which contains a contact-range and pion exchange piece.
  We argue that the contact-range piece survives at leading order,
  while pion exchanges are subleading.
\end{itemize}
This last choice is what defines counting $A$.
Later in Sect.~\ref{sec:EFT-BC} we will revisit the assumptions about coupled
channel dynamics made in the present section and propose other
two countings, $B$ and $C$.
However, for the set of choices explored in this manuscript,
we consistently reproduce the narrow $P_c(4380)$ predicted
in previous works~\cite{Liu:2019tjn,Liu:2019zvb,Guo:2019fdo,Du:2019pij,Du:2021fmf}
instead of the $P_c(4337)$.

\subsection{General considerations}
\label{subsec:general-pc}

We first consider the general features of the lowest order EFT description of
a two-hadron system with a low-lying bound state, which we will
then particularize to the molecular pentaquarks
relevant to this work.
From the EFT point of view the soft scale $Q$ will be given by the momenta
of the hadrons $p$ and the pion mass $m_{\pi}$, to which we will add
the bound state momentum $\gamma_2 = \sqrt{2\mu B_2}$, with $\mu$
the reduced mass of the system and $B_2$ its binding energy.
The hard scale $M$ will be given by the vector meson mass $m_{\rho}$ or
by the typical momentum scale at which the internal structure of
the hadrons becomes apparent.
Contributions to the potential will be categorized as $Q^{\nu}$ (shorthand
for $(Q/M)^{\nu}$), that is, by their power scaling
with respect to the soft scale $Q$.

From counting powers of $Q$ and $M$ directly, the lowest order EFT potential
possible is given by
\begin{eqnarray}
  V^{(0)}(\vec{q}) = V_C(\vec{q}) + V_{\rm OPE}(\vec{q}) \, ,
\end{eqnarray}
with $V_C$ a momentum- and energy-independent contact-range potential and
$V_{\rm OPE}$ the one pion exchange (OPE) potential, which we write as
\begin{eqnarray}
  V_{\rm OPE}(\vec{q}) = - \frac{12 \pi}{\mu\, \Lambda_{\rm OPE}}\,
  \vec{T}_1 \cdot \vec{T}_2\,
  \frac{\vec{S}_{L1} \cdot \vec{q}\,\vec{S}_{L2} \cdot \vec{q}}
       {\vec{q}^2 + m_{\pi}^2} \, , \label{eq:OPE-generic}
\end{eqnarray}
where $\vec{T}_i$ and $\vec{S}_{Li}$ are the isospin and light-spin operators of
hadrons $i=1,2$, $\mu$ the reduced mass of the system and
$\Lambda_{\rm OPE}$ the characteristic OPE scale.
If we count $\Lambda_{\rm OPE}$ as a hard scale, the OPE potential
is of order $Q^0$ and so is the complete lowest order potential.
Yet, it happens that the actual counting of the lowest order potential
can differ from this estimation.

The first obvious modification to the power counting happens
in systems with bound states.
This comes from the observation that the existence of bound states requires
the resummation of a potential $V$, a condition which in terms of
power counting can be written as follows
\begin{eqnarray}
  \mathcal{O} (V) = \mathcal{O} (V G_0 V) \, , \label{eq:loop-count} 
\end{eqnarray}
where $G_0 = 1 / (E_2-H_0)$ is the resolvent operator ($E_2$ is
the center-of-mass energy of the two-body system and $H_0$
its kinetic energy operator), which scales as $Q$
($\sim \gamma_2$) when integrated in a loop:
\begin{eqnarray}
  \int_{\Lambda} \frac{d^3 \vec{q}}{(2 \pi)^3}\,\frac{1}{E_2 - \frac{q^2}{2\mu}}
  = \frac{\mu}{2 \pi}\,\left( \gamma_2 +
  \Lambda\,\beta(\frac{\gamma_2}{\Lambda}) \right) \, , \label{eq:loop}
\end{eqnarray}
where $\Lambda$ represents a cutoff (the loop integral is linearly divergent
and hence requires regularization), $\gamma_2 = \sqrt{- 2 \mu E_2}$ with $\mu$
the reduced mass of the two-body system and $\beta(x)$ a function
that depends on our choice of a regulator.
From this, the potential has to be counted as $V \sim Q^{-1}$
for it to be able to generate a bound state.

When confronting this conclusion with the order $Q^0$ EFT potential, it
is apparent that either the contact- or the finite-range piece
has to be promoted to $Q^{-1}$.
A well-known counting argument states that for a two-body system
with a shallow bound state, the contact-range coupling
will scale as~\cite{vanKolck:1998bw}
\begin{eqnarray}
  c^{(R)} \sim \frac{2\pi}{\mu \sqrt{2\mu B_2}} \sim \mathcal{O}\left(
  \frac{1}{Q} \right) \, , \label{eq:CR-counting}
\end{eqnarray}
where $B_2$ refers to the binding energy of the system and the superscript
${}^{(R)}$ indicates that we are dealing with the {\it renormalized}
coupling ({\it loosely speaking}, the part of the coupling that does
not depend on the cutoff).
This estimation comes from equating $V_C$ and $V_C G_0 V_C$ to make
them comply with the power counting requirements for a bound state
and taking only the finite part of the loop integration of $G_0$
in Eq.~(\ref{eq:loop}).

For the counting of the finite-range piece we first decompose the spin
structure of the OPE potential in a spin-spin and tensor component
\begin{eqnarray}
  V_{\rm OPE}(\vec{q}) &=& - \frac{4 \pi}{\mu\, \Lambda_{\rm OPE}}\,
  \vec{T}_1 \cdot \vec{T}_2\,
  \Big[ \frac{\vec{S}_{L1} \cdot \vec{S}_{L2}\,\vec{q}^2}
    {\vec{q}^2 + m_{\pi}^2} \nonumber \\
    && \qquad \quad  +
    \frac{3\,\vec{S}_{L1} \cdot \vec{q}\,\vec{S}_{L2} \cdot \vec{q} -
    \vec{S}_{L1} \cdot \vec{S}_{L2}\,\vec{q}^2}
       {\vec{q}^2 + m_{\pi}^2} \Big] \, , \label{eq:OPE-decomp}
\end{eqnarray}
which is but a reordering of Eq.~(\ref{eq:OPE-generic}).
The tensor component requires S- to D-wave transitions and
we will assume it to be kinematically suppressed.
Thus, it will not be considered further as part of our calculations.
The spin-spin component acts on S-waves and its counting can be determined
from the calculation of the following ratio 
\begin{eqnarray}
  \frac{\langle V_S G_0 V_S  \rangle}{\langle V_S \rangle}
  &=& \frac{Q}{\Lambda_S} \nonumber \\
  &=& T S\,\frac{m_{\pi}}{\Lambda_{\rm OPE}}\,f(\frac{k}{m_{\pi}}) \, ,
  \label{eq:OPE-counting}
\end{eqnarray}
where $V_S$ refers to the spin-spin component of OPE,
the matrix elements are taken for S-wave scattering
states of center-of-mass momentum $k$ and
$f(x) = 1 - 13/6 x^2 + \mathcal{O}(x^4)$ is a function determining
the momentum dependence of this ratio; its Taylor expansion
is taken from \cite{PavonValderrama:2016lqn} and its usually
decreases with $x$ (meaning that spin-spin OPE becomes
more perturbative at higher momenta), i.e. we can take
$f(k/m_{\pi}) = 1$ without loss of generality.
The first line of Eq.~(\ref{eq:OPE-counting}) is just the generic scaling
of iterated spin-spin OPE when no assumptions are made about
the size of $V_S$, which is encoded in the spin-spin
scale $\Lambda_S$.
The second line is the result of the concrete calculation of the ratio
of the iteration of the potential over the potential, where
$\Lambda_{\rm OPE}$ is the OPE scale as we defined it
in Eq.~(\ref{eq:OPE-generic}) and $S$, $T$ are
simply $\vec{T}_1 \cdot \vec{T}_2$ and
$\vec{S}_{L1} \cdot \vec{S}_{L2}$,
respectively.
From a direct comparison between the two lines, the characteristic
momentum scale associated with spin-spin OPE is
\begin{eqnarray}
  \Lambda_S = \frac{1}{T S} \Lambda_{\rm OPE} \, .
\end{eqnarray}
Depending on its concrete evaluation we will be able to decide
whether spin-spin OPE is $Q^0$ or $Q^{-1}$ within EFT.

\subsection{The $\bar{D}^* \Lambda_c$, $\bar{D} \Sigma_c$ and $\bar{D} \Sigma_c^*$ diagonal potential}

The application of the previous ideas to the $\bar{D}^* \Lambda_c$,
$\bar{D} \Sigma_c$ and $\bar{D} \Sigma_c^*$ systems
is straightforward.
If we assume that these three systems bind or are close to binding,
the ${\rm LO}$ potential is of order $Q^{-1}$ and
only contains a contact-range interaction
\begin{eqnarray}
  V^{(-1)}(\vec{q}, \bar{D}^* \Lambda_c) &=& V_C(\bar{D}^* \Lambda_c) =
  \tilde{d}_a \, ,
  \\
  V^{(-1)}(\vec{q}, \bar{D} \Sigma_c^{(*)}) &=&
  V_C(\bar{D} \Sigma_c^{(*)}) = c_a \, .
\end{eqnarray}
OPE does not contribute though, as neither of these three
systems can exchange one pion (two pions will be the minimum):
OPE is forbidden by isospin symmetry in the $\bar{D}^* \Lambda_c$ system
(and more generally by HQSS in the $\bar{H_c} T_c$ systems),
while for $\bar{D} \Sigma_c$ and $\bar{D} \Sigma_c^*$
the $\bar{D}$ is a pseudoscalar and cannot emit or absorb a pion
(unless it turns into a $\bar{D}^*$, but in this case
we will have a coupled channel effect).

\subsection{Counting of the $\bar{D}^* \Lambda_c$-$\bar{D} \Sigma_c$ and $\bar{D}^* \Lambda_c$-$\bar{D} \Sigma_c^*$ dynamics}
As previously mentioned, the $\bar{D}^* \Lambda_c$ and $\bar{D} \Sigma_c$
thresholds are merely $25\,{\rm MeV}$ away from each other.
This suggests that coupled channel dynamics could be important
in this and other similar cases.

For a two-body system with a shallow bound state, we \linebreak naively expect
coupled channel effects to be suppressed (or enhanced, depending
on the case) by a factor of $B_2 / \Delta_{CC}$, with $B_2$ and
$\Delta_{CC}$ the bound state energy and the mass gap
between the two channels.
Alternatively, in the EFT language, we will compare the characteristic
momentum scales of the bound state and the mass gap,
$Q \sim \gamma_2 = \sqrt{2 \mu B_2}$ and
$M \sim \Lambda_{CC} = \sqrt{2 \mu \Delta_{CC}}$, respectively,
with $\mu$ the reduced mass of the two-body system~\cite{Valderrama:2012jv}.

If we consider the usual, single-channel molecular interpretation of
the $P_c(4312)$ pentaquark --- a $\bar{D} \Sigma_c$ bound state
about $9\,{\rm MeV}$ below threshold -- the EFT expansion parameter
is expected to be the ratio of the two-body binding momentum
(or the pion mass) over the rho meson mass
\begin{eqnarray}
  \frac{Q}{M} \sim \frac{\max{(m_{\pi},\gamma_2)}}{m_{\rho}} \sim 0.18 \, ,
  \label{eq:EFT-expansion-parameter}
\end{eqnarray}
where $\gamma_2 = \sqrt{2 \mu B_2} \sim 137\,{\rm MeV}$ (which coincides
with the pion mass) is the binding momentum, with $\mu$ the reduced mass
of the $\bar{D} \Sigma_c$ or $\bar{D}^* \Lambda_c$ system,
and $m_{\rho} \simeq 770\,{\rm MeV}$ the rho meson mass.
In this case the binding energy is comparable to the $25\,{\rm MeV}$ mass gap
between the $\bar{D}^* \Lambda_c$ and $\bar{D} \Sigma_c$ thresholds,
yielding
\begin{eqnarray}
  {\left( \frac{\gamma_2}{\Lambda_{CC}} \right)}^2 \sim \frac{B_2}{\Delta_{CC}}
  \sim 0.35 \quad
  \mbox{for $\bar{D}^* \Lambda_c$-$\bar{D} \Sigma_c$} \, .
  \label{eq:counting-Pc-CC}
\end{eqnarray}
This ratio, though not particularly large, happens to be larger than
the expansion parameter for the $P_c(4312)$ as a single channel
bound state.
From this, it is apparent that this particular coupled channel effect
enters between ${\rm LO}$ and ${\rm NLO}$ (i.e. next-to-leading-order).
Thus we will simply include the $\bar{D}^* \Lambda_c$-$\bar{D} \Sigma_c$
coupled channel dynamics in the ${\rm LO}$ of the theory.

Naturally, if we consider the $\bar{D}^* \Lambda_c$-$\bar{D} \Sigma_c$
transition potential in the $J = \tfrac{1}{2}$ configuration,
we might consider
the $J=\tfrac{3}{2}$ $\bar{D}^* \Lambda_c$-$\bar{D} \Sigma_c^*$
coupled channel dynamics as well.
The $\bar{D}^* \Lambda_c$-$\bar{D} \Sigma_c^*$ combination offers two possible
interpretations of the $P_c(4337)$ pentaquark:
\begin{itemize}
\item[(i)] The $P_c(4337)$ as a lower mass pole
  in the $\bar{D}^* \Lambda_c$-$\bar{D} \Sigma_c^*$ coupled channel dynamics,
  with the higher mass pole corresponding to the predicted narrow
  $P_c(4380)$ $\bar{D} \Sigma_c^*$ state~\cite{Liu:2019tjn,Liu:2019zvb,Guo:2019fdo,Du:2019pij} (where~\cite{Du:2019pij} claims evidence of its existence within the original data of Ref.~\cite{LHCb:2019kea}).
\item[(ii)] The $P_c(4337)$ as the higher mass pole, which implies that
  this pentaquark is indeed the previously mentioned narrow
  $\bar{D} \Sigma_c^*$ bound state and that its binding
  energy has been underestimated.
\end{itemize}
Actually, if we only use momentum- and energy-independent contact-range
interactions, interpretation (i) can be discarded: with this type of
contact-range interaction a pole originally located below 
$\bar{D}^* \Lambda_c$ can only become a resonance
above this threshold if it hits a second pole further below
the $\bar{D}^* \Lambda_c$ threshold first, where this type of
trajectory is nicely illustrated in Ref.~\cite{Hanhart:2014ssa}.
That is, the ${\rm LO}$ potential requires the lower mass pole to be below
the $\bar{D}^* \Lambda_c$ threshold,
which is incompatible with the assumptions in (i).
Thus, unless we include energy or momentum dependent contact-range interactions
explicitly~\footnote{Here we might be tempted to consider the addition of
  S-to-D-wave interactions (contact-range or pions).
  The rationale is that for $L$-wave interactions with $L \geq 1$,
  when the strength of the interaction weakens, the trajectory of poles
  is such that a bound state becomes a resonance (instead of
  a virtual state as happens in S-waves), check Ref.~\cite{Hanhart:2014ssa}.
  It is not clear whether this strategy will work though. If a non-perturbative
  S-to-D-wave interaction generates a shallow bound state below threshold,
  and this state is mostly S-wave, at low energies the naive expectation
  is that it is possible to describe it in EFT terms with an effective
  S-wave contact-range interaction. Thus, its trajectory in the complex plane
  is naively expected to be that of an S-wave pole,
  i.e. to bounce back at threshold and
  become a virtual state.
  Yet, this is an interesting possibility that deserves further investigation
  in the future.
},
this leaves us with interpretation (ii), which actually might
help explain one feature of the $P_c(4337)$: its large width.
In fact, if the $P_c(4337)$ corresponds to a $\bar{D} \Sigma_c^*$ molecule,
the width of the $\Sigma_c^*$ will have to be added to the intrinsic width of
the $P_c(4337)$ as a pole of the $\bar{D}^* \Lambda_c$-$\bar{D} \Sigma_c^*$
coupled channel dynamics.
It is worth noticing though that the width of a bound $\Sigma_c^*$ will be
narrower than that of a free $\Sigma_c^*$, because of the reduction of
phase space for the pion owing to binding effects.

Independently of the interpretation, the estimations for the convergence
parameter of the EFT expansion for the $P_c(4337)$ as a $\bar{D} \Sigma_c$
molecule and the size of the coupled channel dynamics are
\begin{eqnarray}
  \frac{Q}{M} \sim \frac{\sqrt{2 \mu | E_2 |}}{m_{\rho}} &\sim& 0.42-0.44 \, ,
  \label{eq:counting-Pcprima-diag}
  \\  
  {\left( \frac{\gamma_2}{\Lambda_{CC}} \right)}^2 \sim
  \frac{| E_2 |}{\Delta_{CC}}
  &\sim& 0.54-0.62 \nonumber \\ && \quad
  \mbox{for $\bar{D}^* \Lambda_c$-$\bar{D} \Sigma_c^*$} \, ,
  \label{eq:counting-Pcprima-CC}
\end{eqnarray}
where $E_2$ refers to the center-of-mass energy of the resonance
with respect to either of the thresholds.
That is, the EFT expansion for the $P_c(4337)$ is not expected to work
as well as in the $P_c(4312)$, which will have to be taken into account
when assessing whether these two resonances can be described coherently
within the same formalism.

At this point it is interesting to notice that the previous coupled channel
dynamics are analogous to what might be happening with 
the $P_{cs}(4459)$~\cite{Peng:2020hql}.
The most prosaic molecular explanation of the $P_{cs}(4459)$ pentaquark
is that of a $\bar{D}^* \Xi_c$ bound state with $J=\tfrac{1}{2}$ or
$\tfrac{3}{2}$.
Yet, the $\bar{D} \Xi_c'$ and $\bar{D} \Xi_c^*$ channels are actually
pretty close too, just about $30\,{\rm MeV}$ below and above
the $\bar{D}^* \Xi_c$ threshold.
Repeating the previous arguments we find
\begin{eqnarray}
  \frac{Q}{M} &\sim& \frac{\sqrt{2 \mu | B_2 |}}{m_{\rho}} \sim 0.27 \, ,
  \label{eq:EFT-expansion-parameter-Pcs}
  \\  
  {\left( \frac{\gamma_2}{\Lambda_{CC}} \right)}^2 &\sim&
  \frac{| B_2 |}{\Delta_{CC}}
  \sim 0.54-0.60 \nonumber \\ \quad
  && \mbox{for $\bar{D} \Xi_c$-$\bar{D}^* \Xi_c$ and
    $\bar{D}^* \Xi_c$-$\bar{D} \Xi_c^*$} \, .
  \label{eq:counting-Pcs-CC}
\end{eqnarray}
This prompted us to include this type of coupled channel dynamics,
which breaks the spin degeneracy of the $P_{cs}(4459)$,
in a previous work~\cite{Peng:2020hql}.
Besides, the experimental analysis of Ref.~\cite{Aaij:2020gdg} actually
proposes two possible interpretations for the $P_{cs}(4459)$:
a single peak interpretation, which yields the mass and width
that we previously referred to in Eq.~(\ref{eq:Pcs-single}) or
a double peak interpretation in which the $P_{cs}(4459)$
is actually composed of the following two states
\begin{eqnarray}
  M(P_{cs 1}) &=& 4454.9 \pm 2.7 \, {\rm MeV} \, , \\
  \Gamma(P_{cs 1}) &=& 7.5 \pm 9.7 \, {\rm MeV} \, , \\
  \nonumber \\
  M(P_{cs 2}) &=& 4467.8 \pm 3.7 \, {\rm MeV} \, , \\
  \Gamma(P_{cs 2}) &=& 5.3 \pm 5.3 \, {\rm MeV} \, ,
\end{eqnarray}
which we call $P_{cs 1}$ and $P_{cs 2}$.
The $\bar{D} \Xi_c'$-$\bar{D}^*  \Xi_c$ and $\bar{D}^* \Xi_c$-$\bar{D}  \Xi_c^*$
coupled channel dynamics is able to explain this double peak pattern provided
that the $P_{cs 1}$ and $P_{cs 2}$ are $J=\tfrac{3}{2}$ and $\tfrac{1}{2}$
states, respectively.
From now on (unless stated otherwise),
we will use the two peak solution found by the LHCb experimental
analysis~\cite{Aaij:2020gdg}, i.e. we will assume the existence of
the previous two $P_{cs}$ peaks.

\subsection{The $\bar{D}^* \Lambda_c$-$\bar{D} \Sigma_c$ and $\bar{D}^* \Lambda_c$-$\bar{D} \Sigma_c^*$ transition potential}

The description of the previous coupled channel transitions
depends on the choice of a ${\rm LO}$ EFT potential.
In line with the expected enhancement of contact-range interactions when
there are bound states, we assume that they are included
in the ${\rm LO}$ potential.
OPE also contributes to the $\bar{D}^* \Lambda_c$-$\bar{D} \Sigma_c$ and
$\bar{D}^* \Lambda_c$-$\bar{D} \Sigma_c^*$ transitions, but its
effects will be subleading (we elaborate in a few lines).
From the previous, we end up with the ${\rm LO}$ EFT potential:
\begin{eqnarray}
  V_C(\bar{H}_c T_c - \bar{H_c} S_c) = \lambda^{(O)}\,
  e_b^{(O)} \vec{\sigma}_{L1} \cdot \vec{\epsilon}_{L2} \, ,
\end{eqnarray}
where $S_c = \Sigma_c$, $\Xi_c'$,  $\Omega_c$ or $\Sigma_c^*$, $\Xi_c^*$,
$\Omega_c^*$ are the sextet charmed baryons, $e_b^{(O)}$ is a coupling
(which only involves the octet components of the $\bar{H_c} T_c$ and
$\bar{H}_c S_c$ systems, hence the $^{(O)}$ superscript),
$\lambda^{(O)}$ a numerical flavor factor,
$\vec{\sigma}_{L1}$ the light-spin operator for the light-quark
within the charmed mesons and $\vec{\epsilon}_{L2}$
the polarization vector for the light-diquark
within the sextet baryons.
By particularizing for the $J=\tfrac{1}{2}$
$\bar{D}^* \Lambda_c$-$\bar{D} \Sigma_c$ and $J=\tfrac{3}{2}$
$\bar{D}^* \Lambda_c$-$\bar{D} \Sigma_c^*$ channels,
the potentials happen to be identical and given by
\begin{eqnarray}
  V_C(P_c^N, J=\tfrac{1}{2},\tfrac{3}{2}) &=&
  \begin{pmatrix}
    d_a^{(O)} & -e_b^{(O)} \\
    -e_b^{(O)} & c_a^{(O)} 
  \end{pmatrix} \, , \\
  &=&
  \begin{pmatrix}
    \tilde{d}_a & -\sqrt{3}\,e_b \\
    -\sqrt{3}\,e_b & c_a 
  \end{pmatrix} \, ,
  \label{eq:contact-Pc1}
\end{eqnarray}
where in the second line we have redefined the couplings as to use the same
notation as for the $\bar{D}^* \Lambda_c$ case,
see Eq.(\ref{eq:Vc-DLambdac}).
For the $J=\tfrac{1}{2}$ $\bar{D} \Xi_c'$-$\bar{D}^* \Xi_c$ and
$J=\tfrac{3}{2}$ $\bar{D}^* \Xi_c$-$\bar{D} \Xi_c^*$ systems
we have instead
\begin{eqnarray}
  V_C(P_{cs},J=\tfrac{1}{2}) &=&
  \begin{pmatrix}
    c_a^{(O)} & \frac{e_b^{(O)}}{\sqrt{3}} \\
    \frac{e_b^{(O)}}{\sqrt{3}} & \frac{2}{3}\,d_a^{(S)} + \frac{1}{3}\,d_a^{(O)} 
  \end{pmatrix}
  \\
  &=&
  \begin{pmatrix}
    c_a & e_b \\
    e_b & d_a
  \end{pmatrix} \, , \label{eq:Pcs12}
  \\
  V_C(P_{cs},J=\tfrac{3}{2}) &=&
  \begin{pmatrix}
    \frac{2}{3}\,d_a^{(S)} + \frac{1}{3}\,d_a^{(O)} &
    \frac{e_b^{(O)}}{\sqrt{3}} \\
    \frac{e_b^{(O)}}{\sqrt{3}} & c_a^{(0)} 
  \end{pmatrix} \\
    &=&
  \begin{pmatrix}
    d_a & e_b \\
    e_b & c_a
  \end{pmatrix} \, , \label{eq:Pcs32}
\end{eqnarray}
where, again, we write the potential both in terms of its SU(3) flavor
representations and the couplings we already defined
in Ref.~\cite{Peng:2020hql}.

Regarding the OPE potential, the particular $\bar{H}_c T_c$-$\bar{H_c} S_c$
transition in which it is strongest is
$\bar{D}^{(*)} \Lambda_c$-$\bar{D}^{(*)} \Sigma_c^{(*)}$,
for which it reads~\footnote{
  For the other $\bar{H}_c T_c$-$\bar{H_c} S_c$ transitions, their OPE,
  one-kaon and one-eta exchange potentials can be derived
  from the $\bar{D}^{(*)} \Lambda_c$-$\bar{D}^{(*)} \Sigma_c^{(*)}$ one
  and the relevant SU(3)-flavor symmetry relations.
}
\begin{eqnarray}
  && V_{\rm OPE}(\vec{q}, \bar{D}^{(*)} \Lambda_c \to \bar{D}^{(*)} \Sigma_c^{(*)})
  = \nonumber \\
  && \qquad \quad
  -\frac{g_1 g_3}{\sqrt{2} f_{\pi}^2}\,\vec{\tau}_1 \cdot \vec{t}_2\,
  \frac{\vec{\sigma}_{L1} \cdot \vec{q}\,\vec{\epsilon}_{L2} \cdot \vec{q}}
       {{\vec{q}\,}^2 + m_{\pi}^2} \, ,
\end{eqnarray}
where $g_1$ is the axial coupling of the pion to the charmed mesons
(for which we have $g_1= 0.59 \pm 0.01 \pm 0.07$~\cite{Ahmed:2001xc,Anastassov:2001cw} from the strong $D^*$ decays, or the values extracted
from the updated $D^*$ decay widths~\cite{Zyla:2020zbs},
leading to the estimations
$g_1 = 0.54$~\cite{Mehen:2015efa} or
$g_1 = 0.56 \pm 0.01$~\cite{Yan:2021wdl}),
$|g_3| = 0.973^{+0.019}_{-0.042}$~\cite{Cheng:2015naa} the axial coupling
for the $\Lambda_c \to \Sigma_c^{(*)}$ transition
in the convention by Cho~\cite{Cho:1992cf} (which is related to the convention
by Yan~\cite{Yan:1992gz} by the relation
$g_3^{\rm (Cho)} = -\sqrt{3} g_2^{\rm (Yan)}$;
we notice that Ref.~\cite{Cheng:2015naa} originally uses the Yan convention),
$f_{\pi} = 130\,{\rm MeV}$ the pion weak decay constant and
$m_{\pi} = 138\,{\rm MeV}$ the pion mass.
The $\vec{\tau}_1$ and $\vec{t}_2$ isospin operators are formally analogous to
the $\vec{\sigma}_{L1}$ and $\vec{\epsilon}_{L2}$ light spin operators, and
their evaluation yields $| \vec{\tau}_1 \cdot \vec{t}_2 | = \sqrt{3}$ and $0$
for $I=\tfrac{1}{2}$ and $\tfrac{3}{2}$, respectively.

{
For counting the OPE potential, we rewrite it in the form proposed
in Eq.~(\ref{eq:OPE-decomp})
\begin{eqnarray}
  && V_{\rm OPE}(\vec{q}, \bar{D}^{(*)} \Lambda_c \to \bar{D}^{(*)} \Sigma_c^{(*)})
  = \nonumber \\
  && \qquad -\frac{4 \pi}{\mu\,\Lambda_{\rm OPE}}\,\vec{\tau}_1 \cdot \vec{t}_2\,
  \frac{\vec{\sigma}_{L1} \cdot \vec{\epsilon}_{L2} \, \vec{q}^2}
       {{\vec{q}\,}^2 + m_{\pi}^2} + \dots \, ,
\end{eqnarray}
where the dots indicate the tensor forces,
which we ignore (as they involve D-waves).
From this we find $\Lambda_{\rm OPE} = 1610\,{\rm MeV}$, which
implies a spin-spin OPE scale of $\Lambda_S = 537\,{\rm MeV}$
for $\bar{D}^{(*)} \Lambda_c \to \bar{D}^{(*)} \Sigma_c^{(*)}$
(where $T = \vec{\tau}_1 \cdot \vec{t}_2 = \sqrt{3}$ and
$S = \vec{\sigma}_{L1} \cdot \vec{\epsilon}_{L2} = \sqrt{3})$.
Combined with the coupled channel suppression, the relative size of OPE
with respect to the ${\rm LO}$ diagonal contact-range interaction
happens to be
\begin{eqnarray}
  && \frac{Q}{\Lambda_S}\,{\left( \frac{\gamma_2}{\Lambda_{CC}} \right)}^2
  \sim 0.09 \, , \, 0.14-0.16 \nonumber \\
  && \qquad \qquad \mbox{for $\bar{D}^* \Lambda_c$-$\bar{D} \Sigma_c$,
    $\bar{D}^* \Lambda_c$-$\bar{D} \Sigma_c^*$.} 
\end{eqnarray}
This indicates that spin-spin OPE is ${\rm NLO}$ or smaller.
This type of demotion of the OPE potential is compatible with what happens
in the charmed meson-antimeson~\cite{Valderrama:2012jv} and charmed
baryon-antibaryon systems~\cite{Lu:2017dvm}.
}

{
  At this point it is interesting to compare with~\cite{Du:2021fmf}, a work
  which previously considered the dynamics of
  the $\bar{D}^{(*)} \Lambda_c$ $\to$ $\bar{D}^{(*)} \Sigma_c^{(*)}$ transitions.
  The most important difference with our approach is that
  Ref.~\cite{Du:2021fmf} iterates OPE (both spin-spin and tensor)
  to all orders.
  This requires the inclusion of the S-to-D-wave contacts
  for the numerical renormalization of the amplitudes.
  The reason is that tensor OPE happens to be a singular power-law potential,
  diverging as $1/r^3$ at distances smaller than
  the Compton wave length of the pion~\cite{Beane:2000wh}.
  For hard enough cutoffs this divergence is probed, requiring the inclusion of
  new contact-range couplings for its proper renormalization.
  That is, the pentaquark description of Ref.~\cite{Du:2021fmf} significantly
  differs from ours in what regards to the structure of the contact-range
  potential, where the specific reason why we have less couplings is
  the assumption that tensor OPE is perturbative (as spin-spin OPE
  is known not to modify the power counting of the contacts
  when iterated~\cite{Barford:2002je}).

  Unfortunately, Ref.~\cite{Du:2021fmf} does not explore the question of
  whether tensor OPE is perturbative or not: this will require reexpanding
  the amplitudes in powers of $V_{\rm OPE}$, which is tedious, and
  then comparing this reexpanded amplitude with the original one
  in which OPE is fully iterated.
  However, the fact that the S-to-D-wave contacts are enough to numerically
  renormalize the amplitudes in~\cite{Du:2021fmf} points towards
  the hypothesis that they are perturbative.
  The evidence is circumstantial though and comes from a comparison with
  a previous result regarding the non-perturbative renormalization of
  OPE in the two-nucleon system~\cite{Nogga:2005hy},
  where a tensor contact-range structure (which generates
  the aforementioned S-to-D-wave contacts) was tried,
  but failed to renormalize OPE in all partial waves.
  The reason why the tensor contact-range structure fails is the particular
  way in which singular potentials are renormalized
  in coupled channels~\cite{PavonValderrama:2005wv,PavonValderrama:2005uj},
  yet these results do not preclude the possibility that
  a tensor contact-range potential might work
  in specific cases.
  Be it as it may, the techniques developed in the two-nucleon sector to
  renormalize tensor OPE both perturbatively~\cite{Fleming:1999ee} and
  non-perturbatively~\cite{Nogga:2005hy,Birse:2005um,PavonValderrama:2005wv,PavonValderrama:2005uj}
  could be applied to the case of molecular pentaquarks
  in the future to solve this issue.
}

\subsection{Description of the $P_c(4337)$}

We will now calibrate the ${\rm LO}$ potential as to reproduce the properties
of the $P_c(4312)$, $P_c(4337)$ and the two $P_{cs}$ peaks.
First, a few remarks:
\begin{itemize}
\item[a)] The masses and widths of the $P_c(4312)$ and $P_c(4337)$ depend mostly
  on $c_a$ and $e_b$, respectively, while $\tilde{d}_a$ impacts mostly the
  width of the $P_c(4312)$ (because of the effect of the final
  $\bar{D}^* \Lambda_c$ interaction on the partial decay width
  into this channel).
\item[b)] Conversely, the masses and widths of the $J=\tfrac{1}{2}$ $P_{cs2}$
  depend on $d_a$ and $e_b$, respectively. However this does not
  represent the full width of the $P_{cs1}$ and $P_{cs2}$ pentaquarks:
  it only takes into account the $\bar{D} \Xi_c'$ decay channel
  for $J=\tfrac{1}{2}$, which only {contributes $(1-2)\,{\rm MeV}$
  to its width according to Ref.~\cite{Peng:2020hql}},
  while the $J = \tfrac{3}{2}$ state is predicted to be stable.
  From this and the phenomenological calculation of Ref.~\cite{Xiao:2019gjd},
  which indicates that the $J/\psi \Lambda$ partial decay
  width is also small, we expect the main decay channel
  of the two $P_{cs}$ to be $\bar{D}_s^* \Lambda_c$.
\end{itemize}
With the experimental information available and the assumptions we are making
here, we will be able to determine all the four couplings.

Regarding the widths, we will assume that for molecular pentaquarks they
are saturated by the charmed antimeson - charmed baryon decay channels,
with only a small fraction of the width coming from decays into
charmonium and a light baryon.
This is compatible with the branching ratio limits determined by GlueX~\cite{Ali:2019lzf}, which are
\begin{eqnarray}
  \mathcal{B}(P_c(4312) \to J/\psi p) < 4.6 \% \, ,
\end{eqnarray}
at the $90\%$ confidence level.
For simplicity we will generally ignore the decays involving pions:
concrete calculations~\cite{Burns:2021jlu} show that this type of
partial decay width is very similar in size to the (narrow) decay width of
the charmed baryon within the molecular pentaquark.
Thus, the pion decays will only be important for molecular pentaquarks
containing a $\Sigma_c^*$ charmed baryon.

With the ingredients we have included within our EFT, the only pentaquark
for which we can confidently calculate the width is the $P_c(4312)$:
if assumed to be a $\bar{D} \Sigma_c$ bound state, according
to the arguments in the previous paragraph, its width should
be saturated by the $\bar{D}^* \Lambda_c$ decay channel (the decay
into $\bar{D} \Lambda_c$ is forbidden by HQSS).
For the two $P_{cs}$ pentaquarks, we already mentioned that the main decay
channel is expected to be $\bar{D}_s^* \Lambda_c$ instead.
Finally, for the $P_c(4337)$ the situation is a bit more subtle than
in the previous examples: on the one hand, if it contains a $\Sigma_c^*$,
this will provide a contribution to the width that does not {directly
  appear in our EFT calculations}, which should instead predict
a narrower $P_c(4337)$.
On the other, the momenta involved in the prospective $P_c(4337)$
decays are larger and not necessarily ideal for a ${\rm LO}$ description
in terms of momentum-independent contact-range interactions.
That is, if the $P_c(4337)$ were to really be a $\bar{D} \Sigma_c^*$
molecule, its description would probably require the inclusion of
${\rm NLO}$ contributions if we are to achieve a similar
theoretical accuracy as for the $P_c(4312)$.

From the previous considerations, we will determine $c_a$, $d_a$, $\tilde{d}_a$
and $e_b$ from
\begin{itemize}
\item[a)] the mass and width of the $P_c(4312)$ (as we expect the decay
  width to be saturated by $\bar{D}^* \Lambda_c$),
\item[b)] the masses of the $P_{cs 1}$ and $P_{cs 2}$ pentaquark.
\end{itemize}
We advance that the determination of the couplings
is not unique and there are two types of solutions,
one in which the  $\bar{D}^* \Lambda_c$ diagonal interaction is attractive and
another in which it is repulsive; we will choose the attractive solution,
  as it is better aligned with phenomenological expectations about its sign
  (see the discussion around Eqs.~(\ref{eq:cond-a}-\ref{eq:cond-d})).
For obtaining results we will renormalize the ${\rm LO}$ potential
by including a Gaussian regulator, a cutoff and making the coupling
dependent on the cutoff:
\begin{eqnarray}
  \langle p' | V_C | p \rangle = c(\Lambda)\,
  g(\frac{p'}{\Lambda})\,g(\frac{p}{\Lambda}) \, ,
\end{eqnarray}
with $\Lambda$ the cutoff and $g(x) = e^{-x^2}$ the regulator function.
For the cutoff we will take a central value of $\Lambda = 0.75\,{\rm GeV}$, i.e.
of the order of the rho meson mass, which we will vary
in the $(0.5-1.0)\,{\rm GeV}$ window for estimating
uncertainties.

For convenience we will express $c_a$, $d_a$, $\tilde{d}_a$ and $e_b$ relative
to a reference value, namely the $c_a^{\rm ref}$ coupling that reproduces
the mass of the $P_c(4312)$ as a single-channel $\bar{D} \Sigma_c$
state
\begin{eqnarray}
  c_a^{\rm ref}(\Lambda) = -1.19\,(-(2.17-0.80))\,{\rm fm}^2 \, ,
  \label{eq:Ca-ref}
\end{eqnarray}
where the values in parentheses correspond to
the $(0.5-1.0)\,{\rm GeV}$ cutoff variation.
With this we find:
\begin{eqnarray}
  c_a &=& +1.25\,(1.41-1.18)\,c_a^{\rm ref} \, , \label{eq:Ca-fit1} \\
  \tilde{d}_a &=& +1.32\,(1.40-1.26)\,c_a^{\rm ref} \, , \\
  d_a &=& +1.07\,(1.13-1.04)\,c_a^{\rm ref} \, , \\
  e_b &=& \pm 0.28\,(0.41-0.21)\,c_a^{\rm ref} \label{eq:Eb-fit1} \, ,
\end{eqnarray}
which we will call ``set $A$'' and merits a few comments:
(i) a plus sign indicates an attractive
interaction (the reference coupling is attractive), (ii) for reproducing
the mass of the $P_c(4312)$,  in the coupled channel case $c_a$ has to be
more attractive than in the single channel case to compensate
for the repulsion generated by $e_b$,
(iii) the change in $c_a$ after the inclusion of the coupled channels
($\sim 0.25$) is compatible with the $0.35$ estimation we made for the
relative size of this effect, see Eq.~(\ref{eq:counting-Pc-CC}),
\corr{
  (iv) as in Eq.~(\ref{eq:Ca-ref}),
  the number outside the parentheses are the $\Lambda = 0.75\,{\rm GeV}$
  results, while the first and second number inside the parentheses
  represent the $\Lambda = 0.5$ and $1.0\,{\rm GeV}$ results,
  respectively.
}

We find it surprising that $\tilde{d}_a$ turns out to be so attractive,
which is worth a more extended comment.
First, as already mentioned,
there are actually two possible solutions for the previous
determination of the couplings: one in which $\tilde{d}_a$ is attractive
and another one in which it is repulsive.
However, here we have discarded the repulsive solution because
it will turn out to be incompatible with the width of
the $P_{cs}$ pentaquarks once we include
the $\bar{D}_s^* \Lambda_c$ channel,
as will be explained later.
Second, from phenomenological arguments we expect the following
\begin{eqnarray}
  && \mbox{(a)} \quad c_a, d_a, \tilde{d}_a < 0 \, , \label{eq:cond-a} \\ 
  && \mbox{(b)} \quad c_a \sim d_a \, , \label{eq:cond-b} \\
  && \mbox{(c)} \quad |c_a|, |d_a| > |\tilde{d}_a| \, , \label{eq:cond-c} \\ 
  && \mbox{(d)} \quad |c_a|, |d_a|, |\tilde{d}_a| \gg |e_b| \, .
  \label{eq:cond-d}
\end{eqnarray}
Condition (a) is derived from the observation that the combination of scalar
and vector meson exchange is expected to be attractive for $c_a$, $d_a$ and
$\tilde{d}_a$, though the case for an attractive $\tilde{d}_a$ is weaker
than for $c_a$, $d_a$, see discussion around Eq.~(\ref{eq:VOBE-DLambdac}).
Condition (b) is a consequence of light-meson exchanges, which should
have similar strengths in both cases, a point that seems to be confirmed
in the EFT description of the $P_{cs}$~\cite{Peng:2020hql}.
Condition (c) comes from the observation that vector meson exchange is
repulsive for $\bar{D}^* \Lambda_c$ (while scalar meson exchange
is always attractive).
Condition (d) reflects that $e_b$ has its origins in the magnetic-like coupling
of the vector mesons to the heavy hadrons, which generates a spin-spin
component of the potential that is expected to be weaker
than its central components~\cite{Peng:2020hql,Peng:2020xrf,Peng:2021hkr}.
In this regard, it is interesting to notice
that phenomenological studies of molecular pentaquarks that ignore
these spin-spin interactions do in general a good job
in explaining or even predicting the spectrum~\cite{Xiao:2019gjd,Xiao:2019aya}.
These conditions can be used as {\it priors} on the basis of which to consider
a particular determination of the couplings as being more or less likely.
In particular, Eqs.~(\ref{eq:Ca-fit1}-\ref{eq:Eb-fit1}) fulfill (a), (b), (d),
but not (c), which is the reason why we commented that $\tilde{d}_a$ is
surprisingly attractive. Yet, $\tilde{d}_a$ is probably the coupling
for which our determination should be less reliable.

With these couplings, we predict two bound $\bar{D}^* \Lambda_c$
pentaquarks and the expected $\bar{D} \Sigma_c^*$ pentaquark
\begin{eqnarray}
  M(\bar{D}^* \Lambda_c, \tfrac{1}{2}) &=& 4246.6 \, (4250.1-4244.6)\,
  {\rm MeV} \, , \label{eq:PPLambdac12} \\
  M(\bar{D}^* \Lambda_c, \tfrac{3}{2}) &=& 4257.1 \, (4261.1-4255.0)\,
  {\rm MeV} \, , \label{eq:PPLambdac32} \\
  \nonumber \\
  M(\bar{D} \Sigma_c^*, \tfrac{3}{2}) &=& 4371.2 \, (4370.7-4371.3) \nonumber \\
  &-& i\,5.3 (4.2-5.5) \, {\rm MeV} . \label{eq:PPSigmac32} 
\end{eqnarray}
While the mass of the $\bar{D} \Sigma_c^*$ pentaquark is definitely heavier
than the experimental one, thus reducing the likelihood of 
the $\bar{D}\Sigma_c^*$ interpretation of the $P_c(4337)$,
we nonetheless notice that if we add the $15\,{\rm MeV}$ width of
the $\Sigma_c^*$ to the prediction of the $\bar{D} \Sigma_c^*$ pentaquark,
we will end up with about $25\,{\rm MeV}$, coinciding
with the experimental central value.

However the previous does not take into account a very important difference
between the $J=\tfrac{1}{2}$ and $\tfrac{3}{2}$ channels:
if the $J=\tfrac{3}{2}$ configuration can indeed be identified
with the $P_c(4337)$, the EFT expansion is expected to converge slowly, as shown
in Eqs.~(\ref{eq:counting-Pcprima-diag}) and (\ref{eq:counting-Pcprima-CC}).
That is, a coherent description of the $P_c(4312)$ and $P_c(4337)$ pentaquarks
will benefit from the inclusion of subleading order effects.
Right now, this is not feasible owing to the increase in the number of
parameters that this entails: the combined EFT description of the
$P_c(4312)$ and $P_c(4337)$ contains a total of 6 parameters at
${\rm NLO}$: the three ${\rm LO}$ couplings --- $c_a$, $e_b$ and
$\tilde{d}_a$ --- and the three couplings corresponding to
the $Q^2$ derivative version of the ${\rm LO}$ potential.
In addition, the one pion exchange (OPE) potential is also expected to enter
at ${\rm NLO}$.
We find it worth noticing that not all the $Q^2$ contact-range interactions
enter at ${\rm NLO}$: for instance, there is a tensor coupling between
the S-wave $\bar{D}^* \Lambda_c$ and the D-wave $\bar{D} \Sigma_c$ and
$\bar{D} \Sigma_c^*$ channels, plus a quadrupolar E2-like tensor coupling
between the S- and D-waves of the $\bar{D} \Sigma_c^*$ channel.
These interactions are however only promoted one order with respect to
their naive dimensional estimation and hence enter at ${\rm N^2LO}$.
We refer to Appendix~\ref{app:subleading} for a detailed explanation of
how we count these subleading contact-range interactions.

As the inclusion of ${\rm NLO}$ operators is not a viable strategy
at the moment, this brings us to a different consistency check:
use two different $c_a$'s for the $P_c(4312)$ and $P_c(4337)$ and check
whether their values are consistent within the estimated expansion
parameters in Eqs.~(\ref{eq:counting-Pcprima-diag}) and
(\ref{eq:counting-Pcprima-CC}).
The motivation is that if the power counting of these two molecular candidates
is not the same or it does not converge at the same rate, the couplings are
not necessarily identical if we force the same power counting.
If we perform this exercise with the previously obtained values of $d_a$,
$\tilde{d}_a$ and $e_b$, we get
\begin{eqnarray}
  c_a(P_c') &=& 1.81\,(2.26-1.59)\,{c_a^{\rm ref}} \nonumber \\
  &=& 1.45\,(1.61-1.36)\,{c_a(P_c)} \, ,
\end{eqnarray}
that is, the values of the two couplings differ again by a magnitude
that could be compatible with the unaccounted subleading order
corrections for the $P_c(4337)$, \corr{which we estimated
  to have a relative size of $Q/M \simeq 0.42-0.44$,
  see Eq.~(\ref{eq:counting-Pcprima-diag}).}
In this case the two $\bar{D}^* \Lambda_c$-$\bar{D} \Sigma_c^*$ poles are at
\begin{eqnarray}
  M(\bar{D}^* \Lambda_c, \tfrac{3}{2}) &=& 4250.4\,(4255.3-4247.5)\,{\rm MeV}\, ,\\
  M(\bar{D} \Sigma_c^*) &=& 4337.0 - i\, 8.0\,(6.7-8.7) \, {\rm MeV}\, ,
\end{eqnarray}
where the width of the $P_c(4337)$ is predicted to be $16\,{\rm MeV}$,
which would be compatible with the experimental value once
we add the $\Sigma_c^*$ width.
From this point of view, the previous failure to accurately reproduce
the $P_c(4337)$ pentaquark as a $\bar{D} \Sigma_c^*$ bound state
\corr{would} merely reflect the uncertainty of
the ${\rm LO}$ calculation we are doing here.
But still, the previous conclusion requires that the $P_c(4337)$ is a 
$\bar{D} \Sigma_c^*$ molecule in the first place \corr{(otherwise
  the expansion parameter will be smaller, making it
  impossible to argue for $c_a(P_c)$ and $c_a(P_c')$
  to be compatible)},
which merely show
that this hypothesis is self-consistent \corr{at LO}.
\corr{This potential compatibility between the $P_c(4312)$ and $P_c(4337)$
  could be disproved at ${\rm NLO}$, where the relative difference between
  $c_a(P_c)$ and $c_a(P_c')$ shrinks to $(Q/M)^2 \simeq 0.18-0.19$.
  Yet, there is circumstantial evidence from other
  calculations~\cite{Du:2021fmf,Wang:2019ato} that
  subleading corrections are unlikely to make the descriptions of
  the $P_c(4312)$ and $P_c(4337)$ consistent with each other.
}
Had we used the EFT expansion parameters derived from the $P_c(4312)$
instead, the conclusion would have been different: the $P_c(4337)$ would
have been inconsistent with the $\bar{D} \Sigma_c^*$ molecular
interpretation both at the level of the theoretical and
experimental uncertainties \corr{already at LO,
  as in this case the expansion parameter is 
  $Q/M \sim 0.18$, see Eq.~(\ref{eq:EFT-expansion-parameter}).}

Indeed, this underlines the importance of correctly assessing the uncertainties
of the ${\rm LO}$ calculations for the prospective
$\bar{D} \Sigma_c^*$ pentaquark.
In this regard, in~\cite{Sakai:2019qph}, which includes error estimations,
the mass of the $\bar{D} \Sigma_c^*$ state is calculated to be
$4375.5^{+13.7}_{-23.3}\,{\rm MeV}$ for a cutoff $\Lambda = 1.0\,{\rm GeV}$
and assuming that the $P_c(4440)$ and $P_c(4457)$ are $J=\tfrac{3}{2}$
and $\tfrac{1}{2}$ $\bar{D}^* \Sigma_c$ molecules, respectively.
That is, the calculations of Ref.~\cite{Sakai:2019qph} might be compatible
within errors with the $P_c(4337)$ being a $\bar{D} \Sigma_c^*$ state,
though close to the $2\,\sigma$ level (i.e. a difference of
$39\,{\rm MeV}$ between the central value of this theoretical
prediction and the mass of the $P_c(4337)$, which is to be compared
with an error of $23\,{\rm MeV}$).
\corr{
More recently, Ref.~\cite{Du:2021fmf} includes the $\bar{D}^{(*)} \Lambda_c$
channels, tensor OPE and a series of S-to-D-wave contact-range couplings
required to numerically renormalize the amplitudes (but not the momentum
dependent correction of the $c_a$ coupling), yet it still predicts
the $\bar{D} \Sigma_c^*$ molecule at $4376\,{\rm MeV}$.}
Other calculation that is worth noticing is Ref.~\cite{Wang:2019ato},
which does not include the $\bar{D}^{(*)} \Lambda_c$ channel explicitly
but implicitly as intermediate states in the leading two-pion
exchange (TPE) potential. This effect, which will be subleading order
in our counting (though enhanced with respect to naive estimations
owing to the smaller energy denominators in the TPE diagrams
with $\Sigma_c^{(*)} \to \Lambda_c$ transitions),
generates a $\bar{D} \Sigma_c^*$ potential that is more attractive than
the $\bar{D} \Sigma_c$ one, and predicts the mass of the $\bar{D} \Sigma_c^*$
pentaquark to be about $4362\,{\rm MeV}$, i.e. $10-20\,{\rm MeV}$
lighter than the predictions in theories with contact-interactions
and with/without OPE (e.g. the pionful EFT of Ref.~\cite{Valderrama:2019chc}
estimates the mass of the $\bar{D} \Sigma_c^*$ to be close to
$4380\,{\rm MeV}$).
These and other subleading effects could be worth considering
in the future.

All things considered, the error of the LO calculation seems to be
neither large enough as to easily include the location of
the $P_c(4337)$ nor small enough as to completely
exclude it \corr{(a ${\rm NLO}$ calculation
  would easily solve the issue, though)}.
Here, if besides the cutoff uncertainty we also include the uncertainties
coming (i) from the experimental input, (ii) from HQSS violations (which would
imply that the couplings we have calculated from one state might be off
up to a factor of $\Lambda_{\rm QCD} / m_Q \sim 0.15$ in the charm sector,
with $\Lambda_{\rm QCD} \approx 200\,{\rm MeV}$) and (iii) from
the intrinsic EFT uncertainty (i.e. $Q/M$, for which we will take
the larger $0.27$ estimation for the $P_{cs}(4459)$
in Eq.~(\ref{eq:EFT-expansion-parameter-Pcs})
instead of the smaller $0.18$ for the $P_c(4312)$
in Eq.~(\ref{eq:EFT-expansion-parameter})),
we will arrive at
\begin{eqnarray}
  M(\bar{D} \Sigma_c^*) = 4371.4^{+15.4}_{-17.6} - i\,5.1^{+2.8}_{-4.1} \,{\rm MeV}
  \, ,
\end{eqnarray}
which is compatible (within theoretical uncertainties) with
the aforementioned calculations of
Refs.~\cite{Sakai:2019qph,Wang:2019ato,Valderrama:2019chc},
but not with the mass of
the $P_c(4337)$ (except, again, at the $2\sigma$ level,
and yet only for a pessimistic estimation of
$Q/M$~\footnote{The more conservative choice $Q/M \sim 0.18$ would have yielded
  $M(\bar{D} \Sigma_c^*) = 4371.4^{+13.2}_{-13.0} - i\,5.1^{+2.4}_{-3.4} \,{\rm MeV}$
  instead, leading to a discrepancy above the $2.5\sigma$ level.
}).
If this estimation of the errors is to be considered reliable enough,
it will be improbable that the $P_c(4337)$ is actually
a $\bar{D} \Sigma_c^*$ molecule.
\corr{
  On a different line of thought, if the predicted $P_c(4380)$~\cite{Liu:2019tjn,Liu:2019zvb,Guo:2019fdo,Du:2019pij,Du:2021fmf}
  is experimentally confirmed, the molecular interpretation of the $P_c(4337)$
  as a $\bar{D} \Sigma_c^*$ will also be excluded (unless it is a different
  and unusual experimental manifestation of the $P_c(4380)$).
}

\section{New molecular pentaquarks with antitriplet charmed baryons}

\begin{figure}
  \begin{center}
    \epsfig{figure=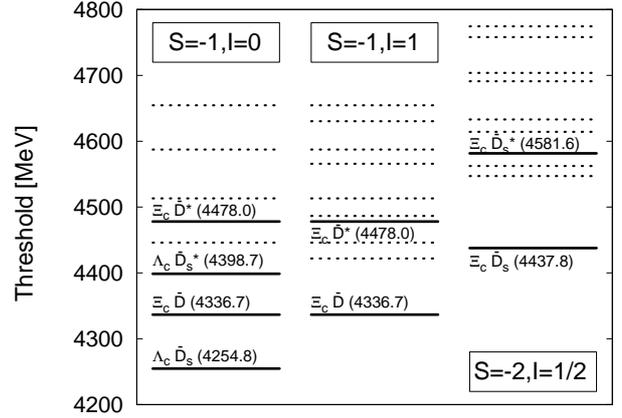,
      width=8.5cm}
\end{center}
\caption{
  Masses of the $\bar{H}_c T_c$ and $\bar{H}_c S_c$ meson-baryon
  thresholds in the strangeness $S=-1$ and $-2$ sectors.
  The thresholds containing an antitriplet charmed baryon are marked using
  solid lines and include a label indicating the explicit meson-baryon
  system under consideration and its mass in ${\rm MeV}$.
  The dotted lines indicate the thresholds containing a sextet charmed baryon,
  but no label is provided to avoid overcluttering the figure:
  in the $S=-1$, $I=0$ sector, the dotted lines correspond to the
  $\bar{D} \Xi_c'$, $\bar{D} \Xi_c^*$, $\bar{D}^* \Xi_c'$ and
  $\bar{D}^* \Xi_c^*$ thresholds (ordered according to increasing mass);
  in the $S=-1$, $I=1$ sector to the $\bar{D}_s \Sigma_c$, $\bar{D} \Xi_c'$,
  $\bar{D}_s \Sigma_c^*$,  $\bar{D} \Xi_c^*$, $\bar{D}_s^* \Sigma_c$,
  $\bar{D}^* \Xi_c'$, $\bar{D}_s^* \Sigma_c^*$ and $\bar{D}^* \Xi_c^*$ thresholds;
  finally, for the $S=-2$, $I=\frac{1}{2}$ sector we have
  $\bar{D}_s \Xi_c'$, $\bar{D} \Omega_c$, $\bar{D}_s \Xi_c^*$,
  $\bar{D} \Omega_c^*$, $\bar{D}_s^* \Xi_c'$, $\bar{D}^* \Omega_c$,
  $\bar{D}_s^* \Xi_c^*$ and $\bar{D}^* \Omega_c^*$, respectively.
  Naively, the relative importance of nearby coupled channels
  is expected to scale as $B_2 / \Delta_{CC}$, with $B_2$ the two-body
  binding energy and $\Delta_{CC}$ the mass gap between thresholds,
  check Eqs.~(\ref{eq:counting-Pc-CC}), (\ref{eq:counting-Pcprima-CC})
  and (\ref{eq:counting-Pcs-CC}) for concrete examples.
}
\label{fig:thresholds}
\end{figure}

It is interesting to notice that the couplings
in set $A$ (i.e. Eqs.~(\ref{eq:Ca-fit1}-\ref{eq:Eb-fit1}))
will lead to the existence of a series of $\bar{H}_c T_c$ bound states.
This can be deduced from the values of the couplings and the single channel
potentials in Eqs.~(\ref{eq:Vc-DsLambdac}-\ref{eq:Vc-DsXic}), leading to
the (single channel) predictions
\begin{eqnarray}
  M(\bar{D}_s \Lambda_c) &=& 4236.1\,(4236.6-4236.3)\, {\rm MeV} \, ,
  \label{eq:SC-P1} \\
  M(\bar{D}^*_s \Lambda_c) &=& 4378.1\,(4379.4-4377.6)\, {\rm MeV} \, , \\
  \nonumber \\
  M(\bar{D} \Xi_c(0)) &=& 4324.2\,(4323.6-4325.1)\, {\rm MeV} \, , \\
  \nonumber \\
  M(\bar{D} \Xi_c(1)) &=& 4310.4\,(4313.0-4309.4)\, {\rm MeV} \, , \\
  M(\bar{D}^* \Xi_c(1)) &=& 4449.4\,(4453.0-4447.4)\, {\rm MeV} \, , \\
  \nonumber \\
  M(\bar{D}_s \Xi_c) &=& 4409.9\,(4413.2-4408.1) {\rm MeV} \, , \\
  M(\bar{D}^*_s \Xi_c) &=& 4551.5\,(4555.8-4548.8)\, {\rm MeV} \, ,
  \label{eq:SC-P7}
\end{eqnarray}
where for the $\bar{D}^{(*)} \Xi_c$ configurations the value in parentheses
represents the isospin ($I=0,1$) and, as usual, the central value
corresponds to $\Lambda = 0.75\,{\rm GeV}$ and the values
in parentheses to $\Lambda = (0.5-1.0)\,{\rm GeV}$.
\corr{However, there are a few instances of nearby $\bar{H}_c T_c$ and
  $\bar{H}_c S_c$ thresholds, as shown in Fig.~\ref{fig:thresholds},
  a fact that points towards the importance of coupled channel dynamics.
  As a consequence}, a consistent prediction of the masses of these states
requires the analysis of the power counting of the different
coupled channel effects relevant to each of these molecules,
as we have already done for the $P_c(4312)$
in Eq.~(\ref{eq:counting-Pc-CC}).

This is done in Table \ref{tab:CC-size} , from which we can appreciate
the existence of a few coupled channel effects
that are worth considering.
For instance, the description of the $J=\tfrac{1}{2}$ $\bar{D}_s^* \Lambda_c$
system will be improved by the addition of the $\bar{D} \Xi_c'$ channel,
leading to the combined $\bar{D}_s^* \Lambda_c$-$\bar{D} \Xi_c'$
potential
\begin{eqnarray}
  V_C(\tilde{P}_{cs}^{\Lambda}, J = \frac{1}{2}) =
  \begin{pmatrix}
    \frac{1}{2}\,(d_a + \tilde{d}_a) & -\sqrt{2}\,e_b \\
    -\sqrt{2}\,e_b & c_a \\
  \end{pmatrix} \, . \label{eq:CC-PLambdatilde}
\end{eqnarray}
Yet, the most clear examples of the importance of coupled channel effects happen
in the $\bar{D}^* \Xi_c(I=1)$ and $\bar{D}_s^* \Xi_c$ molecules, which are really
close to a series of nearby channels:
\begin{itemize}
\item[a)] For the $I=1$, $S=-1$ and $J=\tfrac{1}{2}$ $\bar{D}^* \Xi_c(I=1)$
  system, we will consider the
  $\bar{D}_s \Sigma_c$-$\bar{D} \Xi_c'(1)$-$\bar{D}^* \Xi_c(1)$ basis
  in which the potential reads
  \begin{eqnarray}
    V_C(P_{cs}^{\Sigma} , J=\frac{1}{2}) =
    \begin{pmatrix} \frac{2}{3}\,c_a & -\frac{\sqrt{2}}{3}\,c_a & \sqrt{2}\,e_b \\ -\frac{\sqrt{2}}{3}\,c_a & \frac{1}{3}\,c_a & -e_b \\ \sqrt{2}\,e_b & -e_b & \tilde{d}_a \end{pmatrix} \, . \label{eq:CC-PSigma12}
  \end{eqnarray}
\item[b)] For the the $I=1$, $S=-1$ and $J=\tfrac{3}{2}$
  $\bar{D}^* \Xi_c(I=1)$ system we write the potential in the
  $\bar{D}^* \Xi_c$-$\bar{D}_s \Sigma_c^{*}$-$\bar{D} \Xi_c^*$
  basis:
  \begin{eqnarray}
    V_C(P_{cs}^{\Sigma}, J=\frac{3}{2}) =
    \begin{pmatrix} \tilde{d}_a & \sqrt{2}\,e_b & -e_b \\
        \sqrt{2}\,e_b & \frac{2}{3}\,c_a & -\frac{\sqrt{2}}{3}\,c_a \\
        -e_b & -\frac{\sqrt{2}}{3}\,c_a & \frac{1}{3}\,c_a 
      \end{pmatrix} \, . \label{eq:CC-PSigma32}
  \end{eqnarray}
\item[c)] For the $S=-2$, $J=\tfrac{1}{2}$ $\bar{D}_s^* \Xi_c$ system,
  the basis is $\bar{D}_s \Xi_c'$-$\bar{D} \Omega_c$-$\bar{D}_s^* \Xi_c$
  and the potential:
  \begin{eqnarray}
    V_C(P_{css}^{\Xi}, J=\frac{1}{2}) = \begin{pmatrix} \frac{1}{3}\,c_a & -\frac{\sqrt{2}}{3}\,c_a & e_b \\ -\frac{\sqrt{2}}{3}\,c_a & \frac{2}{3}\,c_a & -\sqrt{2}\,e_b \\ e_b & -\sqrt{2}\,e_b & \tilde{d}_a \end{pmatrix} \, .
    \label{eq:CC-PXi12}
  \end{eqnarray}
\item[d)] Finally, for $S=-2$, $J=\tfrac{3}{2}$ $\bar{D}_s^* \Xi_c$,
  we have $\bar{D}_s^* \Xi_c$-$\bar{D}_s \Xi_c^*$-$\bar{D} \Omega_c^*$ and
  \begin{eqnarray}
    V_C(P_{css}^{\Xi} , J=\frac{3}{2}) =
    \begin{pmatrix} \tilde{d}_a & e_b & -\sqrt{2}\,e_b \\
        e_b & \frac{1}{3}\,c_a & -\frac{\sqrt{2}}{3}\,c_a \\
        -\sqrt{2}\,e_b & -\frac{\sqrt{2}}{3}\,c_a & \frac{2}{3}\,c_a 
    \end{pmatrix} \, . \label{eq:CC-PXi32}
  \end{eqnarray}
\end{itemize}
Here a comment is in order regarding the coupled channels containing
sextet charmed baryons and charmed antimesons, which for $I=1$,
$S=-1$ and $I=\tfrac{1}{2}$, $S=-2$ happen to be an admixture
between octet and decuplet~\cite{Peng:2019wys}, where
\begin{eqnarray}
  | \bar{D}_s^{(*)} \Sigma_c^{(*)} \rangle &=&
  \sqrt{\frac{2}{3}} | 8 \rangle + \sqrt{\frac{1}{3}} | 10 \rangle \, ,
  \label{eq:DsSigma_c} \\
  | \bar{D}^{(*)} \Xi_c^{('/*)} (1) \rangle &=&
  -\sqrt{\frac{1}{3}} | 8 \rangle + \sqrt{\frac{2}{3}} | 10 \rangle \, ,
  \label{eq:DXi_c} \\
  \nonumber \\
  | \bar{D}_s^{(*)} \Xi_c^{('/*)} \rangle &=& 
  \sqrt{\frac{1}{3}} | 8 \rangle + \sqrt{\frac{2}{3}} | 10 \rangle \, ,\\
  | \bar{D}^{(*)} \Omega_c^{(*)} \rangle &=& 
  -\sqrt{\frac{2}{3}} | 8 \rangle + \sqrt{\frac{1}{3}} | 10 \rangle \, ,
\end{eqnarray}
with $| 8 \rangle$ and $| 10 \rangle$ indicating a pure octet or decuplet
state, respectively.
This implies that in principle the potentials
in Eqs.~(\ref{eq:CC-PSigma12}-\ref{eq:CC-PXi32}),
besides the octet coupling $c_a \equiv c_a^{(O)}$,
should also contain a decuplet coupling ($c_a^{(D)}$).
Yet, we have taken $c_a^{(D)} = 0$.

There are two reasons for this choice.
First, for the moment there is no known hidden-charm molecular pentaquark
that is a decuplet, which suggests that there might be less attraction
in this configuration.
This also seems to be supported by phenomenological models~\cite{Liu:2019zvb},
which usually predict less attraction in the decuplets (owing to vector
meson exchange being repulsive in the decuplet case).
If this happens to be the case the addition of a decuplet coupling $c_a^{(D)}$
might be inconsequential after all, as the coupled channel dynamics
will drive the two $\bar{H}_c S_c$ coupled channels to the octet,
which is the minimum energy configuration.

Second, the decuplet configuration can be effectively ignored simply because
the $\bar{H}_c T_c$ systems can only couple to the octet $\bar{H}_c S_c$
configurations, but not to the decuplet ones, which implies that
the coupled channel dynamics driven by the $e_b$ coupling vanishes
if the two $\bar{H}_c S_c$ channels are in a decuplet.
{
This is easy to show by extending the potential for the $P_{cs}^{\Sigma}$
pentaquark as to include the decuplet components, which will generate a
$P_{cs}^{\Sigma^*}$ pentaquark with the quantum numbers of
the $\Sigma^*$ decuplet light baryon, which results in
\begin{eqnarray}
  && V_C(P_{cs}^{\Sigma^{(*)}} , J=\frac{1}{2}) = \nonumber \\
  && \quad  \begin{pmatrix} \frac{2}{3}\,c_a^{(O)} + \frac{1}{3}\,c_a^{(D)} & -\frac{\sqrt{2}}{3}\,(c_a^{(O)} - c_a^{(D)}) & \sqrt{2}\,e_b \\ -\frac{\sqrt{2}}{3}\,(c_a^{(O)} - c_a^{(D)}) & \frac{1}{3}\,c_a^{(O)} + \frac{2}{3}\,c_a^{(D)} & -e_b \\ \sqrt{2}\,e_b & -e_b & \tilde{d}_a \end{pmatrix} \, , \label{eq:CC-PSigma12-full}
\end{eqnarray}
where now we explicitly indicate whether we are dealing with the octet or
decuplet version of the $c_a$ coupling.
Here we can perform a change of basis in which the $\bar{D}_s \Sigma_c$ and
$\bar{D} \Xi_c'(1)$ channels are rewritten in the octet and decuplet basis,
which can be found  by inverting Eqs.~(\ref{eq:DsSigma_c}) and
Eq.~(\ref{eq:DXi_c}).
In this new $| 8 (1) \rangle$-$| 10 (1) \rangle$-$| \bar{D}^* \Xi_c(1) \rangle$
basis (where ``$(1)$'' refers to isospin), the potential reads
\begin{eqnarray}
  \tilde{V}_C(P_{cs}^{\Sigma^{(*)}} , J=\frac{1}{2}) &=&
  \begin{pmatrix} c_a^{(O)} & 0 & -\sqrt{3}\,e_b \\ 0 & c_a^{(D)} & 0 \\ -\sqrt{3}\,e_b & 0 & \tilde{d}_a \end{pmatrix} \, , \label{eq:CC-PSigma12-eigen}
\end{eqnarray}
where it is now evident that the decuplet components
do not mix with the $\bar{D} \Xi_c'(1)$ ($\bar{H}_c T_c$) channel.
This process can be repeated for the $J=\tfrac{3}{2}$ $P_{cs}^{\Sigma^{(*)}}$
configuration, as well as for $J=\tfrac{1}{2}$, $\tfrac{3}{2}$
$P_{cs}^{\Xi^{(*)}}$ ones, with identical outcomes.
Of course, owing to the fact that the thresholds do not have the same mass,
there will be a certain amount of mixing of octet and decuplet.
Yet, we expect this to be a small effect.
}

{
  From the previous two reasons --- namely, that there seems to be no decuplet
  pentaquark and that the decuplet pentaquark configurations decouple
  with the configurations containing an antitriplet
  charmed baryon --- the simplifying assumption
  that $c_a^{(D)} = 0$ seems a sensible choice.
}

After including the coupled channel dynamics suggested by the power counting
estimations of Table \ref{tab:CC-size} (i.e. using the potentials of
Eqs.~(\ref{eq:CC-PLambdatilde}-\ref{eq:CC-PXi32})), we obtain
the predictions of Table \ref{tab:predictions-fit-basic}.
With the couplings in set $A$, all the $\bar{H}_c T_c$ pentaquarks will bind.
The errors shown in Table \ref{tab:CC-size} correspond to varying
the cutoff in the $\Lambda = (0.5-1.0)\,{\rm GeV}$ window and are small,
which simply indicates that the EFT is properly renormalized.
These errors do not reflect the real uncertainty in the location of
$\bar{H}_c T_c$ pentaquarks, as we have not explicitly considered
the propagation of the experimental, HQSS ($\Lambda_{\rm QCD} / m_Q$) and
EFT ($Q/M$) errors.
Instead of including these error sources directly, we will take them
into account indirectly by recalculating the $\bar{H}_c T_c$ pentaquark
spectrum in two other EFTs using different inputs and counting rules.
We will do this in the following section, where we will see that the actual
uncertainty is of the order of tens of ${\rm MeV}$ for most of the states
(as shown in Tables \ref{tab:predictions-fit-intermediate},
\ref{tab:predictions-fit-complete} and \ref{tab:comparison}).

\begin{table*}[t]
  \begin{center}
    \begin{tabular}{|cccccccc|}
      \hline \hline
      Molecule & $I$ & $S$ & $J$ & Expansion parameter &
      \multicolumn{3}{c|}{Nearby thresholds} \\
      \hline \hline
      \multirow{2}{*}{$\bar{D} \Lambda_c$} &
      \multirow{2}{*}{$\frac{1}{2}$} &
      \multirow{2}{*}{$0$} &
      \multirow{2}{*}{$\frac{1}{2}$} &
      \multirow{2}{*}{$0.18$ (${0.29}$)} &
      $\bar{D}^* \Sigma_c$ & $\bar{D}^* \Sigma_c^*$ & $-$   \\
      & & & & & ${0.03}$ (${0.08}$) & ${0.02}$ (${0.07}$) & $-$  \\
      \multirow{2}{*}{$\bar{D}^* \Lambda_c$} &
      \multirow{2}{*}{$\frac{1}{2}$} &
      \multirow{2}{*}{$0$} &
      \multirow{2}{*}{$\frac{1}{2}$} &
      \multirow{2}{*}{$0.18$ (${0.42}$)} &
      $\bar{D} \Sigma_c$ & $\bar{D}^* \Sigma_c^*$ & $-$   \\
      & & & & & ${0.35}$ (${1.88}$) & ${0.04}$ (${0.21}$) & $-$  \\
      \multirow{2}{*}{$\bar{D}^* \Lambda_c$} &
      \multirow{2}{*}{$\frac{1}{2}$} &
      \multirow{2}{*}{$0$} &
      \multirow{2}{*}{$\frac{3}{2}$} &
      \multirow{2}{*}{$0.18$ (${0.37}$)} &
      $\bar{D} \Sigma_c^*$ & $\bar{D}^* \Sigma_c^*$ & $-$   \\
      & & & & & ${0.10}$ (${0.42}$) & ${0.03}$ (${0.16}$) & $-$  \\
      \hline \hline
      \multirow{2}{*}{$\bar{D}_s \Lambda_c$} &
      \multirow{2}{*}{$0$} &
      \multirow{2}{*}{$-1$} &
      \multirow{2}{*}{$\frac{1}{2}$} &
      \multirow{2}{*}{$0.18$ (${0.26}$)} &
      $\bar{D} \Xi_c$ & $\bar{D}^* \Xi_c'$ & $\bar{D}^* \Xi_c^*$   \\
      & & & & & ${0.11}$ (${0.23}$) & ${0.03}$ (${0.06}$) &
      ${0.02}$ (${0.05}$) \\
      \multirow{2}{*}{$\bar{D}_s^* \Lambda_c$} &
      \multirow{2}{*}{$0$} &
      \multirow{2}{*}{$-1$} &
      \multirow{2}{*}{$\frac{1}{2}$} &
      \multirow{2}{*}{$0.18$ (${0.28}$)} &
      $\bar{D} \Xi_c'$ & $\bar{D}^* \Xi_c$ & $\bar{D}^* \Xi_c'$   \\
      & & & & &
      ${0.18}$ (${0.43}$) & ${0.11}$ (${0.26}$) & ${0.05}$ (${0.11}$) \\
      \multirow{2}{*}{$\bar{D}_s^* \Lambda_c$} &
      \multirow{2}{*}{$0$} &
      \multirow{2}{*}{$-1$} &
      \multirow{2}{*}{$\frac{3}{2}$} &
      \multirow{2}{*}{$0.18$ (${0.28}$)} &
      $\bar{D}^* \Xi_c$ & $\bar{D}^* \Xi_c'$ & $\bar{D} \Xi_c^*$   \\
      & & & & &
      ${0.11}$ (${0.26}$) & ${0.05}$ (${0.11}$) & ${0.08}$ (${0.18}$) \\
      \hline \hline
      \multirow{2}{*}{$\bar{D} \Xi_c$} &
      \multirow{2}{*}{$0$} &
      \multirow{2}{*}{$-1$} &
      \multirow{2}{*}{$\frac{1}{2}$} &
      \multirow{2}{*}{$0.18$ (${0.21}$)} &
      $\bar{D}_s \Lambda_c$ & $\bar{D}_s^* \Lambda_c$ & -   \\
      & & & & & $0.11$ ($0.15$) & $0.14$ ($0.20$) & - \\
      \multirow{2}{*}{$\bar{D}^* \Xi_c$} &
      \multirow{2}{*}{$0$} &
      \multirow{2}{*}{$-1$} &
      \multirow{2}{*}{$\frac{1}{2}$} &
      \multirow{2}{*}{$0.18$ (${0.19}$)} &
      $\bar{D} \Xi_c'$ & $\bar{D}_s^* \Lambda_c$ & $\bar{D}_s \Lambda_c$   \\
      & & & & & $0.27$ ($0.32$) & $0.11$ ($0.13$) & $0.04$ ($0.05$) \\
      \multirow{2}{*}{$\bar{D}^* \Xi_c$} &
      \multirow{2}{*}{$0$} &
      \multirow{2}{*}{$-1$} &
      \multirow{2}{*}{$\frac{3}{2}$} &
      \multirow{2}{*}{$0.18$ (${0.24}$)} &
      $\bar{D} \Xi_c^*$ & $\bar{D}_s^* \Lambda_c$ & -   \\
      & & & & & $0.24$ ($0.66$) & $0.11$ ($0.29$) & - \\      
            \hline \hline
      \multirow{2}{*}{$\bar{D} \Xi_c$} &
      \multirow{2}{*}{$1$} &
      \multirow{2}{*}{$-1$} &
      \multirow{2}{*}{$\frac{1}{2}$} &
      \multirow{2}{*}{$0.18$ ($0.31$)} &
      $\bar{D}_s^* \Sigma_c$ & $\bar{D}^* \Xi_c'$ & $\bar{D}_s^* \Sigma_c^*$   \\
      & & & & & $0.04$ ($0.11$) & $0.04$ ($0.10$) & $0.03$ ($0.09$) \\
      \multirow{2}{*}{$\bar{D}^* \Xi_c$} &
      \multirow{2}{*}{$1$} &
      \multirow{2}{*}{$-1$} &
      \multirow{2}{*}{$\frac{1}{2}$} &
      \multirow{2}{*}{$0.18$ ($0.33$)} &
      $\bar{D}_s \Sigma_c$ & $\bar{D} \Xi_c'$ & $\bar{D}_s^* \Sigma_c$   \\
      & & & & & $0.15$ ($0.51$) & $0.27$ ($0.89$) & $0.10$ ($0.33$) \\
      \multirow{2}{*}{$\bar{D}^* \Xi_c$} &
      \multirow{2}{*}{$1$} &
      \multirow{2}{*}{$-1$} &
      \multirow{2}{*}{$\frac{3}{2}$} &
      \multirow{2}{*}{$0.18$ ($0.33$)} &
      $\bar{D}_s \Sigma_c^*$ & $\bar{D} \Xi_c^*$ & $\bar{D}_s^* \Sigma_c$   \\
      & & & & & $1.01$ ($3.36$) & $0.24$ ($0.81$) & $0.10$ ($0.33$) \\
      \hline \hline
      \multirow{2}{*}{$\bar{D}_s \Xi_c$} &
      \multirow{2}{*}{$\frac{1}{2}$} &
      \multirow{2}{*}{$-2$} &
      \multirow{2}{*}{$\frac{1}{2}$} &
      \multirow{2}{*}{$0.18$ (${0.32}$)} &
      $\bar{D}_s^* \Xi_c'$ & $\bar{D}^* \Omega_c$ & $\bar{D}_s^* \Xi_c^*$   \\
      & & & & &
      ${0.03}$ (${0.11}$) & ${0.03}$ (${0.10}$) & ${0.03}$ (${0.09}$) \\
      \multirow{2}{*}{$\bar{D}_s^* \Xi_c$} &
      \multirow{2}{*}{$\frac{1}{2}$} &
      \multirow{2}{*}{$-2$} &
      \multirow{2}{*}{$\frac{1}{2}$} &
      \multirow{2}{*}{$0.18$ (${0.34}$)} &
      $\bar{D}_s \Xi_c'$ & $\bar{D} \Omega_c$ & $\bar{D}^* \Omega_c$   \\
      & & & & & ${0.24}$ (${0.87}$) & ${0.44}$ ($1.57$) & $0.07$ (${0.25}$) \\
      \multirow{2}{*}{$\bar{D}_s^* \Xi_c$} &
      \multirow{2}{*}{$\frac{1}{2}$} &
      \multirow{2}{*}{$-2$} &
      \multirow{2}{*}{$\frac{3}{2}$} &
      \multirow{2}{*}{$0.18$ ($0.34$)} &
      $\bar{D}_s \Xi_c^*$ & $\bar{D} \Omega_c^*$ &  $\bar{D}_s^* \Xi_c'$   \\
      & & & & & ${0.26}$ (${0.92}$) & ${0.16}$ (${0.58}$) & $0.08$ ($0.28$) \\
      \hline
    \end{tabular}
    \caption{Expansion parameter for the EFT description of the $\bar{H}_c T_c$
      pentaquarks and its comparison with the expected size of the coupled
      channel effects of the (two or three) closest thresholds
      with which a given molecular pentaquark can mix.
      ``Molecule'' refers to the two-body system under consideration, $I$,
      $S$, $J$ are its isospin, strangeness and spin, and the expansion
      parameter and relative coupled channel size are calculated
      as in Eqs.~(\ref{eq:EFT-expansion-parameter}) and
      (\ref{eq:counting-Pc-CC}): the first number corresponds to using
      $Q = m_{\pi}$ for the calculation, while the number in parentheses
      uses $Q = \gamma_2$, where the wave number have been
      extracted either from the experimental masses (in the case of
      the $\bar{D}^* \Xi_c$ system), the coupled channel
      calculation of Eqs.~(\ref{eq:PPLambdac12}-\ref{eq:PPSigmac32})
      or the single channel ones of Eqs.~(\ref{eq:SC-P1}-\ref{eq:SC-P7}),
      both of which depend on the couplings in set $A$,
      i.e. Eqs.~(\ref{eq:Ca-fit1}-\ref{eq:Eb-fit1}).
      This determination leads to very attractive couplings, which results
      in a poorer expansion parameter. In contrast, other determinations
      will yield results closer to the $Q = m_{\pi}$ estimations.
      For the masses of the charmed baryons and mesons we use the isospin
      average of the values listed in the Review of Particle
      Physics~\cite{Zyla:2020zbs}.
    }
    \label{tab:CC-size}
  \end{center}
\end{table*}

\begin{table*}[t]
  \begin{center}
    {
    \begin{tabular}{|c|ccc|cc|cc|}
      \hline \hline
      Molecule & $I$ & $S$ & $J$ & Channel(s) & Potential & $M$ &
      $M_{\rm exp}$ \\ 
      \hline \hline
      $\bar{D} \Lambda_c$ & $\frac{1}{2}$ & $0$ & $\frac{1}{2}$ &
      $\bar{D} \Lambda_c$ & $\tilde{d}_a$ &
      {$4129.3^{+1.9}_{-0.3}$} & - \\
      $\bar{D}^* \Lambda_c$ & $\frac{1}{2}$ & $0$ & $\frac{1}{2}$ &
      $\bar{D}^* \Lambda_c$-$\bar{D} \Sigma_c$ &
      Eq.~(\ref{eq:contact-Pc1})
      &
      {$4246.6^{+3.6}_{-1.9}$}
      & - \\
      $\bar{D}^* \Lambda_c$ & $\frac{1}{2}$ & $0$ & $\frac{3}{2}$ &
      $\bar{D}^* \Lambda_c$ -$\bar{D} \Sigma_c^*$ &
      Eq.~(\ref{eq:contact-Pc1})
      &
      {$4257.1^{+4.0}_{-2.1}$} 
      & - \\
      \hline
      \hline
      $\bar{D} \Sigma_c$ & $\frac{1}{2}$ & $0$ & $\frac{1}{2}$ &
      $\bar{D}^* \Lambda_c$-$\bar{D} \Sigma_c$ &
      Eq.~(\ref{eq:contact-Pc1})
      &
      Input(M\&$\Gamma$) & $4311.9 -\frac{i}{2}\,9.8$ \\
      $\bar{D} \Sigma_c^*$ & $\frac{1}{2}$ & $0$ & $\frac{3}{2}$ &
      $\bar{D}^* \Lambda_c$ -$\bar{D} \Sigma_c^*$&
      Eq.~(\ref{eq:contact-Pc1})
      &
      {$4371.4^{+0.1}_{-0.5} - \frac{i}{2}\,10.2^{+0.8}_{-1.8}$}
      & $4337 -\frac{i}{2}\,29$ \\
      \hline \hline
      $\bar{D}_s \Lambda_c$ & $0$ & $-1$ & $\frac{1}{2}$ &
      $\bar{D}_s \Lambda_c$ & $\frac{1}{2}\left( \tilde{d}_a + d_a \right)$ &
      {$4236.1^{+0.5}_{-0.0}$}  & - \\
      $\bar{D}_s^* \Lambda_c$ & $0$ & $-1$ &
      $\frac{1}{2}$ &
      $\bar{D}_s^* \Lambda_c$-$\bar{D} \Xi_c'$ &
      Eq.~(\ref{eq:CC-PLambdatilde}) &
      {$4365.6^{+2.0}_{-1.9}$} & - \\
      $\bar{D}_s^* \Lambda_c$ & $0$ & $-1$ &
      $\frac{3}{2}$ &
      $\bar{D}_s^* \Lambda_c$ & $\frac{1}{2}\left( \tilde{d}_a + d_a \right)$ &
      {$4378.1^{+1.2}_{-0.5}$}
      & - \\
      \hline
      \hline
      $\bar{D} \Xi_c$ & $0$ & $-1$ & $\frac{1}{2}$ &
      $\bar{D} \Xi_c$ & $d_a$ &
      {$4324.2^{+0.9}_{-1.0}$}
      & - \\
      $\bar{D}^* \Xi_c$ & $0$ & $-1$ &
      $\frac{1}{2}$ &
      $\bar{D} \Xi_c'$-$\bar{D}^* \Xi_c$ &
      Eq.~(\ref{eq:Pcs12})
      &
      Input(M){$-\frac{i}{2}\,4.2^{+0.2}_{-0.6}$}
      & $4467.8 - \frac{i}{2}\,5.3$ \\
      $\bar{D}^* \Xi_c$ & $0$ & $-1$ &
      $\frac{3}{2}$ &
      $\bar{D}^* \Xi_c$-$\bar{D} \Xi_c^*$ &
      Eq.~(\ref{eq:Pcs32})
      &
      Input(M) & $4454.9 - \frac{i}{2}\,7.5$ \\
      \hline
      \hline
      $\bar{D} \Xi_c$ & $1$ & $-1$ & $\frac{1}{2}$ &
      $\bar{D} \Xi_c$ & $\tilde{d}_a$ &
      {$4310.4^{+2.6}_{-1.0}$}
      & - \\
      $\bar{D}^* \Xi_c$ & $1$ & $-1$ &
      $\frac{1}{2}$ &
      $\bar{D}_s \Sigma_c$-$\bar{D} \Xi_c'$-$\bar{D}^* \Xi_c$ &
      Eq.~(\ref{eq:CC-PSigma12})
      &
      {$4465.4^{+2.3}_{-1.4} - \frac{i}{2}\,17.0_{-2.7}^{+1.1}$}
      & - \\
      $\bar{D}^* \Xi_c$ & $1$ & $-1$ &
      $\frac{3}{2}$ &
      $\bar{D}^* \Xi_c$-$\bar{D}_s \Sigma_c^{{\ast}}$-$\bar{D} \Xi_c^*$ &
      Eq.~(\ref{eq:CC-PSigma32})
      &
      {$4423.8^{+3.3}_{-3.0}$}
      & - \\
      \hline
      \hline
      $\bar{D}_s \Xi_c$ & $\frac{1}{2}$ & $-2$ & $\frac{1}{2}$ &
      $\bar{D}_s \Xi_c$ & $\tilde{d}_a$ &
      {$4409.9^{+3.3}_{-1.7}$}
      & - \\
      $\bar{D}_s^* \Xi_c$ & $\frac{1}{2}$ & $-2$ &
      $\frac{1}{2}$ &
      $\bar{D}_s \Xi_c'$-$\bar{D} \Omega_c$-$\bar{D}_s^* \Xi_c$ &
      Eq.~(\ref{eq:CC-PXi12})
      &
      {$4572.9^{+2.0}_{-1.6} - \frac{i}{2}\,19.6^{+0.1}_{-1.6}$}
      & - \\
      $\bar{D}_s^* \Xi_c$ & $\frac{1}{2}$ & $-2$ &
      $\frac{3}{2}$ &
      $\bar{D}_s^* \Xi_c$-$\bar{D}_s \Xi_c^*$-$\bar{D} \Omega_c^*$ &
      Eq.~(\ref{eq:CC-PXi32})
      &
      {$4544.2^{+2.8}_{-1.3}$}
      & - \\
      \hline \hline
    \end{tabular}
    }
    \caption{Predictions for the $\bar{H}_c T_c$ family of pentaquarks
      (i.e. pentaquarks containing an antitriplet baryon) and
      the $\bar{D}\Sigma_c^*$ system
      from the couplings in Eqs.~(\ref{eq:Ca-fit1}-\ref{eq:Eb-fit1}),
      which are in turn determined from reproducing the mass and width
      of the $P_c(4312)$ and the masses of the two $P_{cs}$ peaks,
      that is, set $A$ of couplings (for counting $A$).
      ``Channel(s)'' refer to the coupled channel effects included
      in the calculation (which have been chosen according to their
      expected size as estimated in Table \ref{tab:CC-size}),
      ``Potential'' shows the potential in the coupled channel
      space, $M$ is the calculated mass in ${\rm MeV}$ and
      $M_{\rm exp}$ the mass of the experimental candidates,
      if any (also in ${\rm MeV}$).
      The central values are the prediction for $\Lambda = 0.75\,{\rm GeV}$,
      while the errors correspond to varying the cutoff within
      the $(0.5-1.0)\,{\rm GeV}$ range.
    }
      \label{tab:predictions-fit-basic}
  \end{center}
\end{table*}

\section{Alternative power counting schemes} 
\label{sec:EFT-BC}

{
  Power counting depends on the assumptions made about the size of different
  physical effects, assumptions which are in turn constrained
  by the experimental data available.
  For molecular pentaquarks there is no direct experimental data,
  that is, charmed antimeson - charmed baryon scattering data.
  This implies that there is significant freedom on how to organize
  their EFT description, for which the very first assumption
  we are relying on is that a few of the observed
  pentaquarks are indeed molecular.
  Owing to this situation, it will do no harm to revisit our initial
  assumptions (e.g. regarding coupled channel dynamics), propose new power
  counting schemes and recalculate the spectrum of the molecular
  pentaquarks containing an antitriplet charmed baryon.
  Besides power counting $A$, which we have described
  in Sect.~(\ref{sec:EFT-A}), here we will propose
  two additional countings, $B$ and $C$.
}

The power counting estimations of coupled channel effects we have provided
up to this point have relied solely on the relative size of
the propagators in a diagonal and non-diagonal channel.
However, this argument ignores the relative size of the couplings connecting
the channels and this distinction might be important in a few cases,
as we will see.
From a brief inspection of the set $A$ determination of the $c_a$, $d_a$,
$\tilde{d}_a$ and $e_b$ couplings in Eqs.~(\ref{eq:Ca-fit1}-\ref{eq:Eb-fit1})
it is apparent that the size of $e_b$ is smaller
than the other couplings.
This observation might be incorporated into the power counting.

Actually, {there is a phenomenological explanation for this}:
the contact-range couplings follow a multipolar expansion comprised
of central, spin-spin and other terms associated with higher multipolar
light-spin operators, which we can write schematically
as~\cite{Peng:2020xrf}
\begin{eqnarray}
  \langle p' | V_c | p \rangle = f_a +
  f_b\,\vec{S}_{L1} \cdot \vec{S}_{L2} + f_c\,Q_{L1 ij} Q_{L2 ij} + \dots \, ,
  \nonumber \\
\end{eqnarray}
with $f_a$, $f_b$, $f_c$ coupling constants and where $\vec{S}_{L1(2)}$ are
the light-spin operators in the vertex $1(2)$ and $Q_{L1(2) ij}$ are quadrupolar
operators that are defined as $\frac{1}{2} (S_{L1(2) i} S_{L1(2) j} + S_{L1(2) j} S_{L1(2) i}) - \frac{1}{3}\,\delta_{ij} \vec{S}_{L1(2)} \cdot \vec{S}_{L1(2)}$,
while the dots indicate higher order terms.

For the two-hadron systems we have considered, only the first two terms are
non-zero (the quadrupolar term requires the light-spin of both hadrons
to be $S_L \geq 1$ for it to be non-trivial).
It also happens that from phenomenological arguments we expect the terms
in this expansion to decrease in size as we progress
in the multipolar expansion~\cite{Peng:2020hql,Peng:2020xrf}.
That is, the size of the $f_a$ couplings is in general larger
than that of the $f_b$ couplings, as can be appreciated for instance
in Eqs.~(\ref{eq:Ca-fit1}-\ref{eq:Eb-fit1}), i.e. set $A$.
This also explains why molecular pentaquark descriptions
that take $f_b = 0$~\cite{Xiao:2013yca,Xiao:2019gjd,Xiao:2019aya} 
tend to work relatively well.

The previous observation extends to coupled channels: transitions involving
a coupling of the $f_a$ type (or $a$-type) will be enhanced with respect to
those that depend on a coupling of the $f_b$ type (or $b$-type).
For instance, while the $\bar{D}^* \Lambda_c$-$\bar{D} \Sigma_c$ transition
depends on the coupling $e_b$ (check Eq.~(\ref{eq:contact-Pc1})),
this is not true for the $\bar{D}_s \Lambda_c$-$\bar{D} \Xi_c$
transition, which is proportional to the difference of the
$d_a$ and $\tilde{d}_a$ couplings.
In particular, the $\bar{D}_s \Lambda_c$-$\bar{D} \Xi_c$ coupled channel
potential is
\begin{eqnarray}
  && V_C'(P_{cs}^{\Lambda}, I=0,S=-1) = \nonumber \\
  && \qquad
  \begin{pmatrix} \frac{1}{2}(d_a + \tilde{d}_a) & 
    \frac{1}{\sqrt{2}}(d_a - \tilde{d}_a) \\
    \frac{1}{\sqrt{2}}(d_a - \tilde{d}_a) & d_a
  \end{pmatrix} \, , \label{eq:Pc-Lambdac-prima}
\end{eqnarray}
and depending on the specific values of $d_a$ and $\tilde{d}_a$
(and how their difference compares to $e_b$) it could
very well happen that the strength of the $\bar{D}_s \Lambda_c$-$\bar{D} \Xi_c$
dynamics is considerably larger than the naive power counting estimations.
Actually, this is not the case for the couplings of
Eqs.~(\ref{eq:Ca-fit1}-\ref{eq:Eb-fit1}), for which the non-diagonal term
in the previous potential happens to be smaller in size than $e_b$ (but only
if we accept the attractive $\tilde{d}_a$ solution to the fit we made).
Yet, this might be fortuitous, as Eqs.~(\ref{eq:Ca-fit1}-\ref{eq:Eb-fit1})
display relative similar values for $d_a$ and $\tilde{d}_a$.

Owing to the fact that the $(d_a - \tilde{d}_a)/\sqrt{2}$ combination of
couplings might be larger than expected,
it is a good idea to consider this type of
coupled channel dynamics explicitly.
Besides, its dependence on the relative signs of the $d_a$ and
$\tilde{d}_a$ couplings makes this coupled channel particularly
relevant to explore whether the $\bar{D}^{(*)} \Lambda_c$
diagonal interaction is attractive or repulsive.
First, we will do this in a scheme in which we set $e_b = 0$, as we expect
this $b$-type coupling to be smaller in size to the $a$-type couplings.
{This will be power counting $B$.}
From this we will reproduce the mass and width of the single peak solution
of the $P_{cs}(4459)$ pentaquark, i.e. Eq.~(\ref{eq:Pcs-single}),
leading to set $B$:
\begin{eqnarray}
  d_a &=& 1.35\,(1.67-1.23)\,c_a^{\rm ref} \, , \label{eq:Da-fit2} \\
  \tilde{d}_a &=& 0.67\,(0.52-0.74)\,c_a^{\rm ref} \, . \label{eq:Datilde-fit2} 
\end{eqnarray}
These values in turn lead to the predictions of
Table \ref{tab:predictions-fit-intermediate} .

Second, we will consider the case in which all the assumptions behind
power counting $B$ hold, except that now we will take $e_b \neq 0$ again.
We will call this choice power counting $C$.
In this case the $J=\tfrac{1}{2}$ and $\tfrac{3}{2}$ $P_{cs}$ pentaquarks
are $\bar{D}_s^*\Lambda_c$-$\bar{D}\Xi_c'$-$\bar{D}^* \Xi_c$ and
$\bar{D}_s^*\Lambda_c$-$\bar{D}^* \Xi_c$-$\bar{D}\Xi_c^*$ molecules,
respectively, with potentials
\begin{eqnarray}
  && V_C(P_{cs}, J=\frac{1}{2}) = \nonumber \\
  && \quad
  \begin{pmatrix} \frac{1}{2}(d_a + \tilde{d}_a) & -\sqrt{2}\,e_b &
    \frac{1}{\sqrt{2}}(d_a - \tilde{d}_a) \\
    -\sqrt{2}\,e_b & c_a & e_b \\
    \frac{1}{\sqrt{2}}(d_a - \tilde{d}_a) & e_b & d_a \end{pmatrix} \, ,
  \label{eq:Pcs-Lambdac12-full} \\
  && V_C(P_{cs}, J=\frac{3}{2}) = \nonumber \\
  && \quad
  \begin{pmatrix} \frac{1}{2}(d_a + \tilde{d}_a)  &
    \frac{1}{\sqrt{2}}(d_a - \tilde{d}_a) & -\sqrt{2}\,e_b \\
    \frac{1}{\sqrt{2}}(d_a - \tilde{d}_a) & d_a & e_b \\
    -\sqrt{2}\,e_b & e_b & c_a \end{pmatrix} \, .
  \label{eq:Pcs-Lambdac32-full}
\end{eqnarray}
Now we have four parameters that can be determined from different inputs,
including the $P_c(4312)$ and the $P_{cs 1}$ and $P_{cs 2}$.
Here, for a better comparison with the previous calibration (with $e_b = 0$)
we will use the two-peak $P_{cs}(4459)$ solution and determine the four
couplings to the masses and widths of the $P_{c}$ and $P_{cs 1}$ (set $C_1$):
\begin{eqnarray}
  c_a &=& 1.09\,(1.14-1.06)\,c_a^{\rm ref} \, , \label{eq:C1-Ca} \\
  d_a &=& {1.28}\,(1.49-1.19)\,c_a^{\rm ref} \, , \\
  \tilde{d}_a &=& 0.76\,(0.66-0.81)\,c_a^{\rm ref} \, , \\
  e_b &=& \pm 0.14\,(0.20-0.11)\,c_a^{\rm ref} \, , \label{eq:C1-Eb}
\end{eqnarray}
or alternatively with $P_c$ and $P_{cs 2}$ (set $C_2$):
\begin{eqnarray}
  c_a &=& 1.10\,(1.16-1.07)\,c_a^{\rm ref} \, , \label{eq:C2-Ca} \\
  d_a &=& 1.04\,(1.10-1.02)\,c_a^{\rm ref} \, , \label{eq:C2-Da} \\
  \tilde{d}_a &=& 0.79\,(0.71-0.83)\,c_a^{\rm ref} \, , \label{eq:C2-Datilde}\\
  e_b &=& \pm 0.15\,(0.21-0.12)\,c_a^{\rm ref} \, , \label{eq:C2-Eb}
\end{eqnarray}
or directly from $P_{cs 1}$ and $P_{cs 2}$ (set $C_3$):
\begin{eqnarray}
  c_a &=& 1.33\,(1.54-1.23)\,c_a^{\rm ref} \, , \label{eq:C3-Ca} \\
  d_a &=& 1.10\,(1.22-1.06)\,c_a^{\rm ref} \, , \\
  \tilde{d}_a &=& 1.28\,(1.63-1.17)\,c_a^{\rm ref} \, , \\
  e_b &=& \pm 0.35\,(0.56-0.25)\,c_a^{\rm ref} \, , \label{eq:C3-Eb}
\end{eqnarray}
where the couplings are largely compatible with those of set $B$
in Eqs.~(\ref{eq:Da-fit2}) and (\ref{eq:Datilde-fit2}).
In every case, we have chosen the solution with a smaller value of
$|e_b|$ if there is more than one solution (as to represent
a perturbation over counting $B$ and set $B$).
We see that sets $C_1$ and $C_3$ are similar to sets $B$ and $A$, respectively,
while set $C_2$ is the one closest to the phenomenological expectations
that we discussed in Eqs.~(\ref{eq:cond-a}-\ref{eq:cond-d}),
in particular that $c_a \sim d_a$ and $|c_a|, |d_a| > |\tilde{d}_a|$,
which might suggest that set $C_2$ is the most satisfactory of
the three fits.
Thus we will choose set $C_2$ for calculating the spectrum,
which we show in Table \ref{tab:predictions-fit-complete}.
The predictions happen to be qualitatively similar to those of set $A$ and $B$,
Tables \ref{tab:predictions-fit-basic} and
\ref{tab:predictions-fit-intermediate},
but leading to heavier molecular pentaquarks (i.e. to less binding energy) and
to a smaller hyperfine splitting between the $P_{cs1}$ and $P_{cs2}$
pentaquarks than set $A$.

Finally, we compare the $\Lambda = 0.75\,{\rm GeV}$ (central value) predictions
of all the five determinations of the couplings in Table \ref{tab:comparison}.
A few remarks are in order: first, (i) the qualitative characteristics of the
spectrum are similar for all five determinations of the couplings.
Second, (ii) the largest quantitative differences happen for the molecular
configurations that depend on the $\tilde{d}_a$ coupling, which is only
determined indirectly by its effect on the width of $P_c(4312)$ or
$P_{cs1}$ and $P_{cs2}$.
Then, (iii) all this depends on the assumptions we are making in the first
place to determine the couplings, i.e. that the pentaquarks used as input
are all indeed meson-baryon bound states.
It might very well happen that this is not the case, or that the masses,
widths or quantum numbers of the pentaquarks we are using as input might
also change in future experiments.
The bottom-line is that the predictions of $\bar{H}_c T_c$ pentaquarks
we present here are tentative in nature.

\begin{table*}[t]
  \begin{center}
    \begin{tabular}{|cccccccc|}
      \hline \hline
      Molecule & $I$ & $S$ & $J$ & Channel(s) & Potential & $M$ &
      $M_{\rm exp}$ \\ 
      \hline \hline
      $\bar{D} \Lambda_c$ & $\frac{1}{2}$ & $0$ & $\frac{1}{2}$ &
      $\bar{D} \Lambda_c$ & $\tilde{d}_a$ &
      ${(4153.7_{-0.0}^{+0.0})^V}$ & - \\
      $\bar{D}^* \Lambda_c$ & $\frac{1}{2}$ & $0$ &
      $\frac{1}{2}$,$\frac{3}{2}$ &
      $\bar{D}^* \Lambda_c$ &
      $\tilde{d}_a$
      &
      $4295.0^{+0.1}_{-0.0}$ & - \\
      \hline
      \hline
      $\bar{D}_s \Lambda_c$ & $0$ & $-1$ & $\frac{1}{2}$ &
      $\bar{D}_s \Lambda_c$-$\bar{D} \Xi_c$ &
      Eq.~(\ref{eq:Pc-Lambdac-prima})
      &
      $4233.6^{+2.6}_{-6.7}$ & - \\
      $\bar{D}_s^* \Lambda_c$ & $0$ & $-1$ &
      $\frac{1}{2}$,$\frac{3}{2}$ &
      $\bar{D}_s^* \Lambda_c$-$\bar{D}^* \Xi_c$ &
      Eq.~(\ref{eq:Pc-Lambdac-prima})
      &
      $4374.9^{+1.8}_{-6.1}$ &  \\
      \hline
      \hline
      $\bar{D} \Xi_c$ & $0$ & $-1$ & $\frac{1}{2}$ &
      $\bar{D}_s \Lambda_c$-$\bar{D} \Xi_c$ &
      Eq.~(\ref{eq:Pc-Lambdac-prima})
      &
      ${4319.0^{+0.7}_{-0.8} - \frac{i}{2}\,(16.9 \pm {0.2})}$ & - \\
      $\bar{D}^* \Xi_c$ & $0$ & $-1$ &
      $\frac{1}{2}$,$\frac{3}{2}$ &
      $\bar{D}_s^* \Lambda_c$-$\bar{D}^* \Xi_c$ &
      Eq.~(\ref{eq:Pc-Lambdac-prima})
      &
      Input(M\&$\Gamma$) & $4458.8 - \frac{i}{2}\,17.3$ \\
      \hline
      \hline
      $\bar{D} \Xi_c$ & $1$ & $-1$ & $\frac{1}{2}$ &
      $\bar{D} \Xi_c$ & $\tilde{d}_a$ &
      $4336.6^{+0.1}_{-0.0}$ & - \\
      $\bar{D}^* \Xi_c$ & $1$ & $-1$ &
      $\frac{1}{2}$, $\frac{3}{2}$ &
      $\bar{D}^* \Xi_c$ &
      $\tilde{d}_a$
      &
      $4477.8^{+0.2}_{-0.1}$ & - \\
      \hline
      \hline
      $\bar{D}_s \Xi_c$ & $\frac{1}{2}$ & $-2$ & $\frac{1}{2}$ &
      $\bar{D}_s \Xi_c$ & $\tilde{d}_a$ &
      $4437.6^{+0.2}_{-0.1}$ & - \\
      $\bar{D}_s^* \Xi_c$ & $\frac{1}{2}$ & $-2$ &
      $\frac{1}{2}$,$\frac{3}{2}$ &
      $\bar{D}_s^* \Xi_c$ &
      $\tilde{d}_a$
      &
      ${4581.2 ^{+0.4}_{-0.3}}$ & - \\
      \hline \hline
    \end{tabular}
    \caption{Predictions of the $\bar{H}_c T_c$ pentaquarks
      from the fit to the $P_{cs}(4459)$ mass and width
      in the single peak solution,
      i.e. set $B$ in Eqs.~(\ref{eq:Da-fit2}) and (\ref{eq:Datilde-fit2})
      for counting $B$.
      We refer to Table \ref{tab:predictions-fit-basic} for the conventions
      we have used (``Molecule'', $I$, $S$, $J$, etc. as well as
      the cutoff and the uncertainties).
      The only new convention we use is the superscript $^V$, which indicates
      that the pole is a virtual state (instead of a bound state).
    }
    \label{tab:predictions-fit-intermediate}
  \end{center}
\end{table*}

\begin{table*}[t]
  \begin{center}
    {
    \begin{tabular}{|cccccccc|}
      \hline \hline
      Molecule & $I$ & $S$ & $J$ & Channel(s) & Potential & $M$ &
      $M_{\rm exp}$ \\ 
      \hline \hline
      $\bar{D} \Lambda_c$ & $\frac{1}{2}$ & $0$ & $\frac{1}{2}$ &
      $\bar{D} \Lambda_c$ & $\tilde{d}_a$ &
      {$4152.4^{+0.4}_{-0.0}$}
      & - \\
      $\bar{D}^* \Lambda_c$ & $\frac{1}{2}$ & $0$ & $\frac{1}{2}$ &
      $\bar{D}^* \Lambda_c$-$\bar{D} \Sigma_c$ &
      Eq.~(\ref{eq:contact-Pc1})
      &
      {$4286.4^{+0.4}_{-0.0}$} 
      & - \\
      $\bar{D}^* \Lambda_c$ & $\frac{1}{2}$ & $0$ & $\frac{3}{2}$ &
      $\bar{D}^* \Lambda_c$ -$\bar{D} \Sigma_c^*$&
      Eq.~(\ref{eq:contact-Pc1})
      &
      {$4291.4^{+0.4}_{-0.0}$} 
      & - \\
      \hline
      \hline
      $\bar{D} \Sigma_c$ & $\frac{1}{2}$ & $0$ & $\frac{1}{2}$ &
      $\bar{D}^* \Lambda_c$-$\bar{D} \Sigma_c$ &
      Eq.~(\ref{eq:contact-Pc1})
      &
      Input (M \& $\Gamma$) 
      & $4311.9 - \frac{i}{2}\,9.8$ \\
      $\bar{D} \Sigma_c^*$ & $\frac{1}{2}$ & $0$ & $\frac{3}{2}$ &
      $\bar{D}^* \Lambda_c$ -$\bar{D} \Sigma_c^*$&
      Eq.~(\ref{eq:contact-Pc1})
      &
      {$4373.6-\frac{i}{2}\,5.2^{+6.2}_{-2.3}$} 
      & $4337 - \frac{i}{2}\,29$ \\
      \hline \hline
      $\bar{D}_s \Lambda_c$ & $0$ & $-1$ & $\frac{1}{2}$ &
      $\bar{D}_s \Lambda_c$ &
      Eq.~(\ref{eq:Pc-Lambdac-prima})
      &
      {$4248.3^{+0.7}_{-0.9}$} & - \\
      $\bar{D}_s^* \Lambda_c$ & $0$ & $-1$ &
      $\frac{1}{2}$ &
      $\bar{D}_s^* \Lambda_c$-$\bar{D} \Xi_c'$-$\bar{D}^* \Xi_c$ &  
      Eq.~(\ref{eq:Pcs-Lambdac12-full})
      &
      {$4388.5^{+0.2}_{-0.2}$} & - \\
      $\bar{D}_s^* \Lambda_c$ & $0$ & $-1$ &
      $\frac{3}{2}$ &
      $\bar{D}_s^* \Lambda_c$-$\bar{D}^* \Xi_c$-$\bar{D} \Xi_c^*$ &
      Eq.~(\ref{eq:Pcs-Lambdac32-full})
      &
      {$4389.9^{+0.2}_{-0.3}$} & - \\
      \hline
      \hline
      $\bar{D} \Xi_c$ & $0$ & $-1$ & $\frac{1}{2}$ &
      $\bar{D}_s \Lambda_c$-$\bar{D} \Xi_c$ &
      Eq.~(\ref{eq:Pc-Lambdac-prima})
      &
      {$4326.8^{+0.2}_{-0.9} - \frac{i}{2}\,1.0^{+0.1}_{-0.0}$}
      & - \\
      $\bar{D}^* \Xi_c$ & $0$ & $-1$ &
      $\frac{1}{2}$ &
      $\bar{D}_s^* \Lambda_c$-$\bar{D} \Xi_c'$-$\bar{D}^* \Xi_c$ &  
      Eq.~(\ref{eq:Pcs-Lambdac12-full})
      &
      Input(M\&$\Gamma$) 
      & $4467.8 - \frac{i}{2}\,5.3$ \\
      $\bar{D}^* \Xi_c$ & $0$ & $-1$ &
      $\frac{3}{2}$ &
      $\bar{D}_s^* \Lambda_c$-$\bar{D}^* \Xi_c$-$\bar{D} \Xi_c^*$ &
      Eq.~(\ref{eq:Pcs-Lambdac32-full})
      &
      {$4463.9^{+0.7}_{-0.3}-\frac{i}{2}\,0.8^{+0.1}_{-0.0}$}
      & $4454.9 - \frac{i}{2}\,7.5$ \\
      \hline
      \hline
      $\bar{D} \Xi_c$ & $1$ & $-1$ & $\frac{1}{2}$ &
      $\bar{D} \Xi_c$ & $\tilde{d}_a$ &
      {$4334.8^{+0.2}_{-0.0}$}
      & - \\
      $\bar{D}^* \Xi_c$ & $1$ & $-1$ &
      $\frac{1}{2}$ &
      $\bar{D}_s \Sigma_c$-$\bar{D} \Xi_c'$-$\bar{D}^* \Xi_c$ & 
      Eq.~(\ref{eq:CC-PSigma12})
      &
      {$4477.2^{+0.3}_{-0.0} - \frac{i}{2}\,3.1^{+0.1}_{-0.4}$}
      & - \\
      $\bar{D}^* \Xi_c$ & $1$ & $-1$ &
      $\frac{3}{2}$ &
      $\bar{D}^* \Xi_c$-$\bar{D}_s \Sigma_c$-$\bar{D} \Xi_c^*$ & 
      Eq.~(\ref{eq:CC-PSigma32})
      &
      {$4463.8^{+0.7}_{-0.1}$}
      & - \\
      \hline
      \hline
      $\bar{D}_s \Xi_c$ & $\frac{1}{2}$ & $-2$ & $\frac{1}{2}$ &
      $\bar{D}_s \Xi_c$ & $\tilde{d}_a$ &
      {$4435.4^{+0.4}_{-0.0}$}
      & - \\
      $\bar{D}_s^* \Xi_c$ & $\frac{1}{2}$ & $-2$ &
      $\frac{1}{2}$ &
      $\bar{D}_s \Xi_c'$-$\bar{D} \Omega_c$-$\bar{D}_s^* \Xi_c$ &
      Eq.~(\ref{eq:CC-PXi12})
      &
      {$4582.2^{+0.1}_{-0.0} -\frac{i}{2}\,6.1^{+0.1}_{-0.5}$}
      & - \\
      $\bar{D}_s^* \Xi_c$ & $\frac{1}{2}$ & $-2$ &
      $\frac{3}{2}$ &
      $\bar{D}_s^* \Xi_c$-$\bar{D}_s \Xi_c^*$-$\bar{D} \Omega_c^*$ &
      Eq.~(\ref{eq:CC-PXi32})
      &
      {$4577.3^{+0.2}_{-0.0}$}
      & - \\
      \hline \hline
    \end{tabular}
    }
    \caption{
      Predictions of the $\bar{H}_c T_c$ pentaquarks
      (plus the $\bar{D} \Sigma_c$ and $\bar{D} \Sigma_c^*$ systems)
      from the fit to the $P_{c}(4312)$ and $P_{cs 2}$ masses
      and widths, i.e. set $C_2$ in Eqs.~(\ref{eq:C2-Ca}-\ref{eq:C2-Eb})
      for counting $C$.
      We refer to Table \ref{tab:predictions-fit-basic} for the conventions
      we have used (``Molecule'', $I$, $S$, $J$, etc. as well as
      the cutoff and the uncertainties).
    }
    \label{tab:predictions-fit-complete}
  \end{center}
\end{table*}

\begin{table*}[t]
  \begin{center}
    \begin{tabular}{|ccccccccc|}
      \hline \hline
      Molecule & $I$ & $S$ & $J$ & Set A & Set B & Set $C_1$ & Set $C_2$ & Set $C_3$\\ 
      \hline \hline
      $\bar{D} \Lambda_c$ & $\frac{1}{2}$ & $0$ & $\frac{1}{2}$ &
      $4129.3$ & ${(4153.7)^V}$ & $4153.0$ & $4152.4$ & $4131.4$ \\
      $\bar{D}^* \Lambda_c$ & $\frac{1}{2}$ & $0$ &
      $\frac{1}{2}$ &
      $4246.6$ & $4295.0$ & $4288.2$ & $4286.4$ & $4238.1$ \\
      $\bar{D}^* \Lambda_c$ & $\frac{1}{2}$ & $0$ &
      $\frac{3}{2}$ &
      $4257.1$ & $4295.0$ & $4292.5$ & $4291.4$ & $4252.9$ \\
      \hline \hline
      $\bar{D} \Sigma_c$ & $\frac{1}{2}$ & $0$ & $\frac{1}{2}$ &
      Input(M\&$\Gamma$) & - & Input(M\&$\Gamma$) & Input(M\&$\Gamma$) & $4313.6$ \\
      $\bar{D} \Sigma_c^*$ & $\frac{3}{2}$ & $0$ & $\frac{1}{2}$ &
      $4371.4$ &  - & $4373.6$ & $4373.6$ & $4371.2$  \\
      \hline \hline
      $\bar{D}_s \Lambda_c$ & $0$ & $-1$ & $\frac{1}{2}$ &
      $4236.1$ & $4233.6$ & $4238.2$ & $4248.3$ & $4235.5$ \\
      $\bar{D}_s^* \Lambda_c$ & $0$ & $-1$ &
      $\frac{1}{2}$ &
      $4365.6$ & $4374.9$ & $4378.5$ & $4388.5$ & $4354.4$ \\
      $\bar{D}_s^* \Lambda_c$ & $0$ & $-1$ &
      $\frac{3}{2}$ &
      $4378.1$ & $4374.9$ & $4379.2$ & $4389.9$ & $4365.2$ \\
      \hline
      \hline
      $\bar{D} \Xi_c$ & $0$ & $-1$ & $\frac{1}{2}$ &
      $4324.2$ & ${4319.0}$ & $4318.5$ & $4326.8$ & $4323.4$ \\
      $\bar{D}^* \Xi_c$ & $0$ & $-1$ &
      $\frac{1}{2}$ &
      Input(M) & Input(M\&$\Gamma$) & $4460.2$ & Input(M\&$\Gamma$) &
      Input(M\&$\Gamma$) \\
      $\bar{D}^* \Xi_c$ & $0$ & $-1$ &
      $\frac{3}{2}$ &
      Input(M) & Input(M\&$\Gamma$) & Input(M\&$\Gamma$) & $4463.9$ &
      Input(M\&$\Gamma$) \\
      \hline
      \hline
      $\bar{D} \Xi_c$ & $1$ & $-1$ & $\frac{1}{2}$ &
      $4310.4$ & $4336.6$ & $4335.5$ & $4334.8$ & $4312.5$ \\
      $\bar{D}^* \Xi_c$ & $1$ & $-1$ &
      $\frac{1}{2}$ &
      $4465.4$ & $4477.8$ & $4477.6$ & $4477.2$ & $4471.6$\\
      $\bar{D}^* \Xi_c$ & $1$ & $-1$ &
      $\frac{3}{2}$ &
      $4423.8$ & $4477.8$ & $4465.6$ & $4463.8$ & $4414.5$ \\
      \hline
      \hline
      $\bar{D}_s \Xi_c$ & $\frac{1}{2}$ & $-2$ & $\frac{1}{2}$ &
      $4409.9$ & $4437.6$ & $4436.2$ & $4435.4$ & $4412.0$  \\
      $\bar{D}_s^* \Xi_c$ & $\frac{1}{2}$ & $-2$ &
      $\frac{1}{2}$ &
      $4572.9$ & ${4581.2}$ & $4582.6$ & $4582.2$ & $4580.4$ \\
      $\bar{D}_s^* \Xi_c$ & $\frac{1}{2}$ & $-2$ &
      $\frac{3}{2}$ &
      $4544.2$ & ${4581.2}$ & $4578.5$ & $4577.3$ & $4551.4$ \\
      \hline \hline
    \end{tabular}
    \caption{Comparison of the predictions of the $\bar{H}_c T_c$ pentaquarks
      (plus the $\bar{D} \Sigma_c$ and $\bar{D} \Sigma_c^*$ systems)
      from the five determinations of the couplings we have considered,
      i.e. set
      $A$ (Eqs.~(\ref{eq:Ca-fit1}-\ref{eq:Eb-fit1})),
      $B$ (Eqs.~(\ref{eq:Da-fit2}-\ref{eq:Datilde-fit2})),
      $C_1$ (Eqs.~(\ref{eq:C1-Ca}-\ref{eq:C1-Eb})),
      $C_2$ (Eqs.~(\ref{eq:C2-Ca}-\ref{eq:C2-Eb})) and
      $C_3$ (Eqs.~(\ref{eq:C3-Ca}-\ref{eq:C3-Eb})).
      We refer to Table \ref{tab:predictions-fit-basic} for the conventions
      we have used (``Molecule'', $I$, $S$, $J$, etc.).
      As for the cutoff, we take $\Lambda = 0.75\,{\rm GeV}$.
      The differences among the predictions for each set could be interpreted
      as the expected theoretical uncertainty for the masses of
      these pentaquarks.
    }
    \label{tab:comparison}
  \end{center}
\end{table*}

\section{Conclusions}
From a theoretical point of view, the new $P_c(4337)$ pentaquark is more
puzzling than the previous $P_c(4312)$, $P_c(4440)$, $P_c(4457)$
and $P_{cs}(4459)$ pentaquarks.
While the later are easily explained as $\bar{D} \Sigma_c$,
$\bar{D}^* \Sigma_c$ and $\bar{D}^* \Xi_c$ bound states
(which does not necessarily mean they are),
the new $P_c(4337)$ does not fit into the previous pattern so nicely.
From HQSS we expect the existence of a $\bar{D} \Sigma_c^*$ bound state,
yet its mass is usually predicted to lie in the $4370-4380\,{\rm MeV}$
range, about $30-40\,{\rm MeV}$ above the mass of the $P_c(4337)$.
Thus the $P_c(4337)$ offers a more interesting challenge than
the other pentaquarks.

Here we explore a few explanations of the new $P_c(4337)$: (i) that it is
a $\chi_{c0} p$ bound state, (ii) that, owing to the Breit-Wigner
parametrization being not suitable to near threshold states,
the $P_c(4337)$ is actually a $\bar{D} \Sigma_c$ molecule
while the $P_c(4312)$ is a $\bar{D}^* \Lambda_c$ molecule instead
and (iii) that the $P_c(4312)$ and \linebreak
$P_c(4337)$ are actually a consequence
of the $\bar{D}^* \Lambda_c$-$\bar{D} \Sigma_c$ and
$\bar{D}^* \Lambda_c$-$\bar{D} \Sigma_c^*$ coupled
channel dynamics.

The first explanation, which requires $J^P = \tfrac{1}{2}^+$ \linebreak
for the $P_c(4337)$, is theoretically plausible and
has the advantage of leaving the usual molecular understanding of the
\linebreak $P_c(4312)$, $P_c(4440)$, $P_c(4457)$ and $P_{cs}(4459)$ pentaquarks
unchanged, as the new $P_c(4337)$ will be unrelated to the previous pentaquarks.
If this is the case, in analogy to the results of Ref.~\cite{Eides:2017xnt}
for $\psi' p$, other $\chi_{c0} B_8$, $\chi_{c1} B_8$ and $\chi_{c2} B_8$
pentaquarks are to be expected, with $B_8$ representing baryons
in the light baryon octet.
This explanation has been recently explored in Ref.~\cite{Ferretti:2021zis},
for instance (though in this reference the $P_c(4337)$ happens to be related
to the $P_{cs}(4459)$, which is not the case in the present work).

The second explanation is grounded in the observation that the difference
in the masses of the $P_c(4312)$ and $P_c(4337)$ states coincides
with the mass gap of the $\bar{D}^* \Lambda_c$ and
$\bar{D} \Sigma_c$ thresholds.
It implies $J^P = \tfrac{1}{2}^-$ for the $P_c(4337)$, while suggesting
a $J^P = \tfrac{1}{2}^-$ and $\tfrac{3}{2}^-$ double
peak nature of the $P_c(4312)$ (i.e. the two spin states of
$\bar{D}^* \Lambda_c$), which is an interesting possibility.
Besides, it will imply a strong attractive interaction for all
the $\bar{H}_c T_c = \bar{D}^{(*)} \Lambda_c$, $\bar{D}_s^{(*)} \Lambda_c$,
$\bar{D}^{(*)} \Xi_c$ and $\bar{D}_s^{(*)} \Xi_c$ systems and
thus the existence of a few new molecular pentaquarks.
For checking this possibility it will be necessary to adapt the techniques
of Refs.~\cite{Albaladejo:2015lob,Fernandez-Ramirez:2019koa,Yang:2020nrt}
to the particular case of the previous two channels and pentaquarks.

The third explanation hopes to reproduce the $P_c(4312)$ and $P_c(4337)$
pentaquarks as poles in the $\bar{D}^* \Lambda_c$-$\bar{D} \Sigma_c$ and
$\bar{D}^* \Lambda_c$-$\bar{D} \Sigma_c^*$ coupled channel systems.
This requires the $P_c(4312)$ and \linebreak $P_c(4337)$ to have
$J^P = \tfrac{1}{2}^-$ and $\tfrac{3}{2}^-$, respectively.
Though this explanation fails to reproduce the location of the $P_c(4337)$
accurately, it nonetheless shows that the $P_c(4312)$ and the double peak
interpretation of the $P_{cs}(4459)$ can be described with the same set
of parameters in the molecular picture.
Actually, the aforementioned failure might be attributable to a very prosaic
explanation: the non-relativistic EFT we are using to describe
the $P_c(4312)$ and $P_c(4337)$ states might converge better
in the former than in the later.
In fact, the LO uncertainty in the mass of the $P_c(4337)$ as a
$\bar{D} \Sigma_c^*$ molecule could be compatible with
its experimental location, at least for pessimistic
estimations of these uncertainties.
This is not what we could call a very exciting conclusion, but
it is not a particularly plausible interpretation either once
we consider more sober estimations of the LO errors.
Indeed, for most calculations the preferred mass of the $\bar{D} \Sigma_c^*$
molecule is consistently centered around $(4370-4380)\,{\rm MeV}$,
in agreement with previous theoretical works~\cite{Liu:2019tjn,Liu:2019zvb,Guo:2019fdo,Du:2019pij,Du:2021fmf} (and in disagreement
with the hypothesis of the $P_c(4337)$ being a $\bar{D} \Sigma_c^*$
molecule).
Nonetheless it is still true that the
$\bar{D}^* \Lambda_c$-$\bar{D} \Sigma_c^*$ explanation points again
towards the existence of a few unobserved molecular pentaquarks
involving the antitriplet charmed baryons.
Using different power countings for the coupled channel effects and inputs,
we have calculated a series of possible predictions for the masses of
these pentaquarks.

{
The EFT description of the $P_c(4312)$ and $P_c(4337)$ as
$\bar{D}^* \Lambda_c$-$\bar{D} \Sigma_c$ and
$\bar{D}^* \Lambda_c$-$\bar{D} \Sigma_c^*$ systems
depends on how we count coupled channel effects.
Within EFT, the inclusion of coupled channel dynamics depends on two factors,
the mass difference between the two channels and the relative size of
the transition potential.
The first factor is straightforward: a larger mass difference
between thresholds implies a larger suppression of a given
coupled channel effect.
For the second factor, the transition potential,
we have considered three power countings, $A$, $B$ and $C$:
$A$ assumes that the antitriplet-sextet charmed
baryon transitions (e.g. $\bar{D}^* \Lambda_c$-$\bar{D} \Sigma_c$)
are enhanced with respect to naive expectations, $B$ in contrast
chooses to enhance the antitriplet-antitriplet transitions
(e.g. $\bar{D}_s^* \Lambda_c$-$\bar{D}^* \Xi_c$),
while $C$ enhances both.
While the mass of the $\bar{D} \Sigma_c^*$ molecule is fairly
independent of which choice we have made among the previous countings,
this is not the case for all the meson-baryon systems
we have considered here.
For instance, owing to the closeness of the thresholds involved,
the predictions for the mass of the $I=1$, $J=\tfrac{3}{2}$
$\bar{D}^* \Xi_c$-$\bar{D}_s \Sigma_c^*$-$\bar{D} \Xi_c^*$
bound state vary by a few tens of ${\rm MeV}$ depending
on the coupled channel dynamics and hence,
on which counting we use.
Finally, the impact of pion exchanges (particularly tensor OPE) remains a very
interesting subject for future investigations: though we suspect that
pion exchanges might be perturbative, this is yet to be proven with explicit
comparisons between perturbative and non-perturbative calculations.
}

\section*{Acknowledgments}

We thank Feng-Kun Guo for comments and suggestions.
M.P.V. would also like to thank the IJCLab of Orsay, where part of
this work has been done, for its long-term hospitality.
This work is partly supported by the National Natural Science Foundation
of China under Grants No. 11735003, No. 11835015, No. 11975041, No. 12047503
and No. 12125507,  the Chinese Academy of Sciences under Grant No. XDPB15,
the Fundamental Research Funds for the Central Universities and
the Thousand Talents Plan for Young Professionals.

\appendix
\section{Subleading order contributions to the molecular pentaquark potential}
\label{app:subleading}

In this appendix we briefly review the subleading order potential
for the anticharmed meson and charmed baryon system.
If we consider the contact-range potential first, the operators
with up to two derivatives (i.e. containing up to ${\vec{q}\,}^2$ terms)
that are pertinent to the molecular pentaquarks
we are considering here are
\begin{eqnarray}
  V_C(\vec{q}, \bar{H}_c S_c) &=&
  \hat{c}_{a} + \hat{c}_{a2}\, {\vec{q}\,}^2 \nonumber \\
  &+& \left( \hat{c}_b + \hat{c}_{b2}\,{\vec{q}\,}^2 \right)\,
  \vec{\sigma}_{L1} \cdot \vec{S}_{L2} \nonumber \\
  &+& \hat{c}_{c2} \left(
  \vec{\sigma}_{L1} \cdot \vec{q} \, \vec{S}_{L2} \cdot \vec{q} -
  \frac{1}{3}\,\vec{\sigma}_{L1} \cdot \vec{S}_{L2}\,{\vec{q}\,}^2
  \right) \nonumber \\
  &+& \hat{c}_{d2}\,Q_{L2 ij}\,q_i\,q_j \, , \\
  V_C(\vec{q}, \bar{H}_c T_c) &=& \hat{d}_{a} + \hat{d}_{a2}\, {\vec{q}\,}^2
  \, , \\
  V_C(\vec{q}, \bar{H}_c T_c - \bar{H_c} S_c) &=&
  \left( \hat{e}_{b} + \hat{e}_{b2}\,{\vec{q}\,}^2 \right)\,
  \vec{\sigma}_{L1} \cdot \vec{\epsilon}_{L2} \nonumber \\
  &+& \hat{e}_{c2}\,\left(
  \vec{\sigma}_{L1} \cdot \vec{q} \, \vec{\epsilon}_{L2} \cdot \vec{q} -
  \frac{1}{3}\,\vec{\sigma}_{L1} \cdot \vec{\epsilon}_{L2}\,{\vec{q}\,}^2
  \right) \, , \nonumber \\
\end{eqnarray}
where by pertinent we refer to S-wave or S-to-D-wave operators (in addition
to these there will be P-wave operators too, but they are unrelated to
the S-wave bound states we are dealing with here).
In the expressions above $\hat{c}$, $\hat{d}$ and $\hat{e}$ are short-hand
for their SU(3)-flavor decomposition, which we have not written down
explicitly, $\vec{q}$ is the momentum exchanged
between the hadrons, $\vec{\sigma}_{L1}$ are the Pauli matrices
as applied to the light-spin of the $\bar{D}$ and $\bar{D}^*$ mesons,
$\vec{\epsilon}_{L2}$ and $\vec{S}_{L2}$ the polarization vector and
the spin operator of the light diquark within the sextet charmed baryons and
$Q_{L2 ij}$ a quadrupolar light-spin operator defined as
\begin{eqnarray}
  Q_{L2 ij} = \frac{1}{2} (S_{L2 i} S_{L2 j} + S_{L2 j} S_{L2 i}) - \frac{1}{3} \vec{S}_{L2}^2 \delta_{ij} \, .
\end{eqnarray}

Each of these contacts work as follows:
\begin{itemize}
\item[a)] The $\hat{c}_a$, $\hat{d}_a$, $\hat{e}_b$ contacts and their
  derivative counterparts ($\hat{c}_{a2}$, $\hat{d}_{a2}$, $\hat{e}_{b2}$)
  only act on S-waves.

\item[b)] This is also the case for $\hat{c}_b$ and $\hat{c}_{b2}$, which
  generate the hyperfine splitting among the different spin configurations of
  the $\bar{D}^* \Sigma_c$ and $\bar{D}^* \Sigma_c^*$ systems.
  However, we do not consider these systems here.

\item[c)] The $\hat{c}_{c2}$ contact generates S-to-D-wave mixing in the
  $\bar{D}^* \Sigma_c$ and $\bar{D}^* \Sigma_c^*$ systems
  (and the transitions of the previous two systems to and from
  the $\bar{D} \Sigma_c$ and $\bar{D} \Sigma_c^*$).
  Thus $\hat{c}_{c2}$ is not of direct relevance for this work either.

\item[d)] The $\hat{c}_{d2}$ contact also provides S-to-D-wave mixing,
  but it affects the $\bar{D} \Sigma_c^*$ and $\bar{D}^* \Sigma_c^*$
  systems instead, i.e. systems containing a $\Sigma_c^*$ (and
  the transitions to and from $\bar{D} \Sigma_c$ and
  $\bar{D}^* \Sigma_c$).
  It will generate a bit of attraction in $\bar{D} \Sigma_c^*$,
  though at subleading orders only.
  
\item[e)] Finally, the $\hat{e}_{c2}$ contact represents an S-to-D-wave
  transition between $\bar{D}^* \Lambda_c$ and $\bar{D} \Sigma_c$,
  $\bar{D} \Sigma_c^*$, etc.
\end{itemize}
All this means that the couplings directly relevant for the molecular
pentaquarks we are considering here are the octet piece of
$\hat{c}_a$, $\hat{c}_{a2}$ and $\hat{c}_{d2}$,
the singlet and octet pieces of $\hat{d}_a$ and $\hat{d}_{a2}$ and
the octet piece of $\hat{e}_b$, $\hat{e}_{b2}$ and $\hat{e}_{c2}$,
meaning a total of 10 independent couplings
with less than two derivatives.
The previous number assumes that the power counting is determined by naive
dimensional analysis (NDA), i.e. terms with no derivatives count as $Q^0$ and
terms with two derivatives as $Q^2$.

However,
{
  if we are dealing with non-perturbative physics,
  the previous assumptions about the power counting do not apply,
  as already discussed in Sect.~\ref{subsec:general-pc}
  for the ${\rm LO}$ couplings.
  Here, we will briefly explain the power counting of the subleading couplings
  from a Wilsonian renormalization group (RG) point of
  view~\cite{Birse:1998dk,PavonValderrama:2014zeq,Valderrama:2016koj}.
  That is, we will consider the evolution of the contact-range couplings
  as we soften the cutoff under the constraint that observable
  quantities remain cutoff independent.
  If we have a theory with soft and hard momentum scales $Q$ and $M$,
  where $M \gg Q$, and use a momentum space cutoff $\Lambda$ to regularize
  the loop integrals, first we will begin by assuming that
  when the cutoff is of the order of the hard scale, $\Lambda \sim M$,
  the contact-range couplings are natural
  in $M$~\cite{Valderrama:2016koj,Epelbaum:2017byx}
  \begin{eqnarray}
    c_d(\Lambda \sim M) \propto \frac{1}{M^{2+d}} \, , 
  \end{eqnarray}
  where $d$ is the number of derivatives of the contact-range operator
  that multiplies the coupling $c_d$.
  As the cutoff evolves towards $\Lambda \sim Q$, the contributions that
  originally were in the loops will be reabsorbed in the couplings
  themselves, leading to a change in their value
    \begin{eqnarray}
      c_d(\Lambda \sim Q) \propto \frac{1}{M^{2+d}}\,{\left(\frac{M}{Q}\right)}^a
      \, ,
    \end{eqnarray}
  where we have assumed a power-law evolution for simplicity.
  If $a > 0$, this coupling will be enhanced at low energies by a factor
  of $1/Q^a$ in the counting; this usually happens in the presence of
  non-perturbative physics, which depend on loops.
  Hence, as loop contributions are reduced as the cutoff softens,
  the couplings have to compensate by increasing their contribution.
  Of course, it is important to notice that here the cutoff is being used
  as an {\it analysis tool}: we reduce $\Lambda$ to uncover which
  couplings are more relevant at low energies.
  Its use is different than in standard calculations, for which $\Lambda > Q$
  is necessary to avoid finite cutoff effects at low energies.
  We notice that the standard expectation of no divergences at large $\Lambda$
  for a renormalized calculation is fulfilled within the Wilsonian approach.
  By virtue of the RG evolution being derived from the cutoff independence of
  observables, and provided we have included all couplings required at the
  order we are calculating, there should be cutoff independence (modulo
  higher order corrections) when the cutoff is
  increased to $\Lambda \sim M$.
}

{
  In this picture, if we reconsider the $P_c(4312)$ and $P_{cs}(4459)$ as
  $\bar{D} \Sigma_c$ and $\bar{D}^* \Xi_c$ shallow bound states,
  respectively, we end up with
\begin{eqnarray}
  c_a(\Lambda), d_a(\Lambda) &\propto& \frac{2 \pi}{\mu}\,\frac{1}{\Lambda}
  \, ,
\end{eqnarray}
with $\mu$ the reduced mass of any of these two systems and
$\Lambda$ the cutoff, where the previous expression
can be derived from Eqs.~(\ref{eq:loop-count}) and
(\ref{eq:loop}) and taking the binding energy
equal to zero for simplicity.
In contrast, in Sect.~\ref{subsec:general-pc} we discussed instead
the renormalized couplings, for which we basically ignored
the cutoff in Eqs.~(\ref{eq:loop-count}) and
(\ref{eq:loop}), leading to Eq.~(\ref{eq:CR-counting}).
Yet, the end result in terms of power counting is the same, because
as we reduce the cutoff to $\Lambda \sim Q$, we find
\begin{eqnarray}
  c_a, d_a &\sim& Q^{-1} \, ,
\end{eqnarray}
to which we also have also added
\begin{eqnarray}
  \tilde{d}_a, e_b &\sim& Q^{-1} \, ,
\end{eqnarray}
where the promotion of $\tilde{d}_a$ implicitly assumes a bound or virtual
state in the $\bar{D} \Lambda_c$ system.
Meanwhile, the promotion of $e_b$ is not strictly required but will
nonetheless be useful if we want to describe the hyperfine
splitting of the $P_{cs1}$ and $P_{cs2}$ pentaquarks at ${\rm LO}$.
This leave us with 4 couplings at ${\rm LO}$ (or $Q^{-1})$.
Despite this picture being very different from the more standard one that
we laid out in Sect.~\ref{subsec:general-pc}, they nonetheless arrive to
the same conclusion.
}

For ${\rm NLO}$ (or $Q^0$) we have to take into account that the counting
of the contacts involving two derivatives is affected by
the ${\rm LO}$ wave functions.
In particular they will be promoted, as we can check by imposing
renormalization group invariance to these couplings.
The idea is that if we have a subleading order contact-range interaction
of the form
\begin{eqnarray}
  \delta V_C = \delta c(\Lambda)\,\times\,\mathcal{O}_{12}(\vec{q}^2) \, ,
\end{eqnarray}
where $\delta c(\Lambda)$ is a subleading order coupling and $\mathcal{O}_{12}$
a contact-range operator involving at least two derivatives, then the running
of $\delta c(\Lambda)$ can be determined from the condition~\cite{PavonValderrama:2014zeq}
\begin{eqnarray}
  \frac{d}{d \Lambda}\,
  \langle \Psi_{\rm LO}' |\, \delta V_{C} | \Psi_{\rm LO} \rangle = 0 \, ,
\end{eqnarray}
where $| \Psi_{\rm LO} \rangle$ and $| \Psi_{\rm LO}' \rangle$ refer to the
initial and final ${\rm LO}$ wave functions.
That is, the matrix elements of $\delta V_C$ should be cutoff independent.
Usually, the evaluation of this matrix element leads to an equation of the type
\begin{eqnarray}
  \frac{d}{d \Lambda}\,\left[ \Lambda^{a}\,\delta c(\Lambda) \right] = 0 \, , 
\end{eqnarray}
which implies an enhancement of $\delta c(\Lambda)$ by a factor of $1/Q^a$
as we evolve it from $\Lambda \sim M$ to $\Lambda \sim Q$.
The enhancement is easy to deduce from the expected behavior of the S- and
D-wave wave functions~\cite{Valderrama:2016koj}, yielding:
\begin{itemize}
\item[a)] $1/Q^1$ ($a=1$) for the S-to-D-wave couplings
  ($\hat{c}_{d2}$, $\hat{e}_{c2}$) and
\item[b)] $1/Q^2$ ($a=2$) for the S-wave couplings
  ($\hat{c}_{a2}$, $\hat{d}_{a2}$, $\hat{e}_{b2}$).
\end{itemize}
As a consequence, the S-wave couplings are promoted from $Q^2$ to $Q^0$
(or ${\rm NLO}$), while all the tensor and quadrupolar operators
are promoted from $Q^2$ to $Q^1$ (or ${\rm N^2LO}$).
This leave us with only 4 new couplings (or a total of 8) at ${\rm NLO}$ and
2 more (or a total of 10) at ${\rm N^2LO}$.
{
  The enhancement we find here for the S- and S-to-D-wave derivative
  interactions is identical to the one previously calculated
  in the two-nucleon system when OPE is either
  perturbative~\cite{Kaplan:1998tg,Kaplan:1998we}
  or not present~\cite{Chen:1999tn}, despite
  the obvious differences in the methodology.
}


%

\end{document}